\documentclass[onecolumn,notitlepage,letterpaper,superscriptaddress,nofootinbib,longbibliography,prd]{revtex4-2}

\usepackage[english]{babel}
\usepackage[T1]{fontenc}
\usepackage[utf8]{inputenc}
\usepackage{amsmath,accents}
\usepackage{amsfonts,color}
\usepackage{amssymb,float}
\usepackage{mathptmx}
\usepackage{comment}
\usepackage{mathrsfs}

\usepackage{graphicx}
\usepackage{subfigure}
\usepackage[dvipsnames]{xcolor}

\usepackage[pdftoolbar = true, pdfstartview = FitH, pdfmenubar = true]{hyperref}
\hypersetup{
    colorlinks=true,
    linkcolor=blue,
    filecolor=magenta,      
   citecolor=blue
}

\newcommand{\weff}{\mathrm{w}_{\mathrm{eff}}}

\newcommand{\Veff}{V_{\mathrm{eff}}}
\newcommand{\meff}{m_{\mathrm{eff}}}
\newcommand{\IM}{\mathcal{I}_{\mathrm{m}}}
\newcommand{\IV}{\mathcal{I}_{{V}}}
\newcommand{\IP}{\mathcal{I}_{\Phi}}
\newcommand{\Acal}{\mathcal{A}}
\newcommand{\Bcal}{\mathcal{B}}

\newcommand{\MP}{M_{\mathrm{Pl}}}
\newcommand{\DD}{\mathrm{d}}

\begin{document}

\title{Global Portraits of Nonminimal Inflation: Metric and Palatini}

\author{Laur Järv} 
\email{laur.jarv@ut.ee}
\affiliation{Institute of Physics, University of Tartu, W.\ Ostwaldi 1, 50411 Tartu, Estonia}
\author{Sotirios Karamitsos}
\email{skaramitsos@phys.uoa.gr}
\affiliation{Department of Physics, University of Athens, Zographou 157 84, Greece}
\author{Margus Saal}
\email{margus.saal@ut.ee}
\affiliation{Institute of Physics, University of Tartu, W.\ Ostwaldi 1, 50411 Tartu, Estonia}



\begin{abstract}
In this paper, we study the global phase space dynamics of single nonminimally coupled scalar field inflation models in the metric and Palatini formalisms. Working in the Jordan frame, we derive the scalar-tensor general field equations and flat FLRW cosmological equations, and present the Palatini and metric equations in a common framework. We show that inflation is characterized by a ``master'' trajectory from a saddle-type de Sitter fixed point to a stable node fixed point, approximated by slow roll conditions (presented for the first time in the Palatini formalism). We show that, despite different underlying equations, the fixed point structure and properties of many models are congruent in metric and Palatini, which explains their qualitative similarities and their suitability for driving inflation. On the other hand, the global phase portraits reveal how even models which predict the same values for observable perturbations differ, both to the extent of the phase space physically available to their trajectories, as well as their past asymptotic states. We also note how the slow roll conditions tend to underestimate the end of inflationary accelerated expansion experienced by the true nonlinear ``master'' solution. The explicit examples we consider range from the metric and Palatini induced gravity quintic potential with a Coleman-Weinberg correction factor, to Starobinsky, metric, and Palatini nonminimal Higgs, as well as second-order pole and several nontrivial Palatini models.
\end{abstract}

\maketitle

\section{Introduction}
The inflationary paradigm is well established as a natural explanation for the observed isotropy and homogeneity of the universe on large scales, as well as the flatness of the universe and the absence of exotic relics \cite{Ratra:1987rm, Mukhanov:1990me, Lyth:1998xn}. However, its predictive power comes to the forefront in calculations of the spectrum of density perturbations. During the accelerated expansion of the Universe, the background evolution of the inflaton field drives the generation of primordial fluctuations. These fluctuations result in inhomogeneities, which, in turn, give rise to the nearly scale-invariant spectrum observed in the cosmic microwave background (CMB) and large-scale structure. Observation of this spectrum is essential in testing the inflationary hypothesis, as inflation is predicted to leave a noticeable imprint on the primordial perturbations.

An abundance of inflationary models has been proposed in recent years \cite{Martin:2013tda}. Many of those have not survived the increasingly tightening observational constraints of Planck \cite{Planck:2018vyg}, which has severely constrained the values of the spectral tilt $n_s$ and tensor-to-scalar ratio~$r$. As such, the simpler candidates for the inflationary theory (such as monomial models) have been ruled out, forcing us to consider more sophisticated alternatives.  A popular approach to inflationary theory is to introduce a coupling function between the inflaton and the scalar curvature \cite{Accetta:1985du, Lucchin:1985ip, Bezrukov:2007ep}, giving rise to a broader class of gravity models referred to as scalar-tensor theories~\cite{Capozziello:2011et} within the general framework of modified gravity~\cite{Nojiri:2017ncd}. Such a coupling function may be motivated by quantum corrections to the low-energy effective action that emerges when high-energy degrees of freedom are integrated out. Trading the minimal coupling for a nonminimal one introduces new phenomenological features, including phases of superaccelerated expansion \cite{Gunzig:2000kk, Jarv:2009zf}. Despite the presence of more involved equations, the spectrum of perturbations for nonminimal models can still be predicted relatively easily by adopting a frame-invariant approach \cite{Jarv:2016sow, Kuusk:2016rso, Burns:2016ric}.

In the standard metric approach to gravity, in which inflation was originally formulated, the notion of parallel transport as defined by the connection is firmly linked to the notion of distance through the metric-dependent Christoffel symbols.
However, in differential geometry, the connection is in principle
an independent quantity from the metric, and the curvature invariants are directly the functions of the connection,
rather than the metric. 
In the spirit of this notion, in Palatini general relativity we instead write the Ricci scalar in the Einstein-Hilbert action as a function of the general connection, which we do not assume to take on a particular form. Then, the connection field equation forces it to reduce to the usual Levi-Civita form, making the two formalisms physically equivalent. However, for extensions of general relativity like with nonminimally coupled scalar fields the Palatini approach delivers equations that are physically different from the usual metric approach \cite{Lindstrom:1975ry, Lindstrom:1976pq,vandenBergh:1981, Burton:1997pe}, leading to theories with different predictions and opening up an entire class of new theories \cite{Olmo:2011uz}. In particular, applying this principle to inflation \cite{Bauer:2008zj} gives ample novel possibilities for model building~\cite{Tenkanen:2020dge, Gialamas:2023flv}. The algorithm for quick computation of the observables can be extended to the Palatini case as well \cite{Jarv:2020qqm}.

The theoretical study of either metric or Palatini inflationary models, leaving aside intricacies such as reheating which occurs after the end of inflation, usually proceeds by considering the evolution of the universe in slow-roll solutions. It is now understood that slow-roll behaves as an attractor for a very robust set of initial conditions in most well-behaved theories of inflation \cite{Belinsky:1985zd, Linde:1985ub, Liddle:1994dx}. 
However, restricting ourselves to the study of attractor solutions or small deviations thereof runs the danger of obfuscating the global features and details of the model. In particular, models with the same slow-roll predictions can very well have divergent global behaviors, making them particularly interesting targets for study.

The methods of dynamical systems make it possible to construct phase space diagrams that present information about the behavior of all solutions of a system. Such diagrams have found prolific use in the study of the late universe dominated by dark energy \cite{Bahamonde:2017ize}, and have also been used in the study of inflationary scalar fields from early on \cite{Belinsky:1985zd, Halliwell:1986ja, Amendola:1990nn, Burd:1991ns, Capozziello:1993xn, Kolitch:1994qa, Cornish:1995mf, Copeland:1997et, deOliveira:1997jt, Holden:1998qg, Gunzig:2000kk, Gunzig:2000yj, Gunzig:2000ce, Saa:2000ik, Heard:2002dr, Felder:2002jk, Carloni:2007eu, Abdelwahab:2007jp, Jarv:2008eb, Hrycyna:2008gk, Szydlowski:2013sma, Hrycyna:2013yia, Remmen:2013eja, Arefeva:2012sqa, Skugoreva:2014gka}, but only occasionally. It was realized only relatively recently by Alho and Uggla that the phenomenon of the inflationary attractor solution owes its properties to being a heteroclinic orbit from a de Sitter kind fixed point with saddle-type features to the final stable node in the phase space \cite{Alho:2014fha, Alho:2017opd}, although this has been noted (qualitatively at least) earlier as well \cite{Belinsky:1985zd, Felder:2002jk, Urena-Lopez:2007zal, Urena-Lopez:2011gxx}. Such orbits originate from a primordial point with de Sitter effective barotropic index, which typically resides in the asymptotics of the phase space \cite{Alvarez:2019vbp, Quiros:2020bcg, Hrycyna:2020jmw, Jarv:2021qpp, Jarv:2021ehj, Hrycyna:2021yad, Hrycyna:2022der, Alho:2023pkl} (although not always \cite{Jarv:2021qpp}). This becomes evident only when one looks at the \textit{global} picture of the dynamics. The aforementioned primordial state is rather special, as almost all other solutions begin at a different asymptotic state characterized by a power law expansion of the scale factor \cite{Felder:2002jk, Carloni:2007eu, Sami:2012uh, Skugoreva:2014gka, Jarv:2021qpp}, and approach the inflationary orbit only later. A closely related issue is the relationship between the inflationary orbit, which is a solution of the full nonlinear equations, and the curve expected from the slow roll conditions, which is a solution of the approximated equations. It turns out that the slow roll curve starts at the primordial inflationary point tangentially to the inflationary orbit \cite{Alho:2014fha, Alho:2017opd, Alho:2023pkl, Jarv:2021qpp}, but later gradually deviates and finally tends to underestimate the end phase of inflationary expansion \cite{Urena-Lopez:2007zal, Grain:2017dqa, Jarv:2021qpp}.

Drawing a clear distinction between the different asymptotic states, as well as having an undistorted depiction of the evolution from the beginning until the end of inflation and beyond, depends heavily on the choice of suitable dynamical variables. There are various options available, differing in mathematical practicality as well as ease of physical interpretation. In this paper, we adopt the variables and methods developed in Refs.\ \cite{Jarv:2021qpp, Jarv:2021ehj} and extend them from the metric to the Palatini formulation of single-field models that feature a generic nonminimal coupling to gravity, noncanonical kinetic terms, and arbitrary nonnegative potentials. As the models in the Palatini formalism can be recast as standard scalar-tensor theories, we may expect that the qualitative picture of inflation ruled by a heteroclinic orbit still holds. However, the precise details need to be worked out and illustrated by explicit examples. In particular, since rather different metric and Palatini models can give identical predictions for the spectral tilt and tensor-to-scalar ratio \cite{Jarv:2016sow, Jarv:2020qqm}, it would be interesting to gain a deeper understanding of this phenomenon from the global phase space perspective, and see which dynamical features there are that could break the observational degeneracy.

The structure of the present paper is as follows. In Sections \ref{sec:scaltens} and \ref{sec:FLRWcosmology}, we review scalar-tensor theories of gravitation in the metric and Palatini formalism, as well as their application to cosmology, presenting them in a common framework. In Section~\ref{sec:observables} we outline how to extract the inflationary observables from such theories, and then present the dynamical systems approach to inflation in Section~\ref{sec:dynsys}. We illustrate these general investigations by first examining in detail the induced gravity model of quintic potential with the Coleman-Weinberg correction factor, which predicts identical observables in the metric and Palatini cases, in Sec.\ \ref{sec:identicalactionsidenticalobs}. Then, we turn to various different actions which nevertheless predict the same observables in Section \ref{sec:differentactionsidenticalobs}: these include the Starobinsky model and nonminimal Higgs model in the metric formalism, a second-order pole model in minimal coupling in which metric and Palatini cases are equivalent, and three models in the Palatini formalism, viz.\ a nonminimally coupled nontrivial potential model, a nonminimally coupled nontrivial kinetic term model, and an induced gravity inverse power law potential model. Later, in Sec.\ \ref{sec:identicalactionsdifferentobs}, we complement the previous section by looking at the Palatini nonminimal Higgs and metric induced gravity inverse power law models, which give different observables than their counterparts in the other formalism. We finish by presenting our conclusions and outlook in Section \ref{sec:conclusions}.

Throughout this paper, we adopt the following conventions: the metric signature is $(-+++)$, and we use a natural unit system where the speed of light $c=1$ and the universal gravitational constant $G=1$, meaning that all dimensions are encoded in powers of the reduced Planck mass $\MP$.

\section{Scalar-tensor gravity in metric and Palatini formalisms}
\label{sec:scaltens}

The aim of this section is to present the metric and Palatini formulations of scalar-tensor gravity in a unified framework where the respective field equations can be compared term by term. Let us begin by writing down a generic Jordan frame action functional for a class of theories which include scalar self-interaction only through the scalar field but not its derivatives as follows \cite{Flanagan:2004bz, Jarv:2014hma, Jarv:2015kga, Kozak:2018vlp}:\footnote{A more general Palatini action can also contain additional terms due to the nonmetricity of the connection \cite{Kozak:2018vlp}. However, it is possible to show that these extra terms can be eliminated by a conformal transformation of the independent connection, see Appendix A of Ref. \cite{Jarv:2020qqm}.}
\begin{equation} \label{action:flanagan:1}
S = \frac{1}{2}\int_{V_4} \mathrm{d}^4x\sqrt{-g}\left\lbrace {\mathcal A}(\Phi) {\mathcal{R}} - {\mathcal B}(\Phi)g^{\mu\nu}\partial_\mu\Phi \partial_\nu\Phi - 2 \, \mathcal{V}(\Phi)\right\rbrace 
+ S_{(\mathrm{mat})}\left[e^{\sigma(\Phi)} g_{\mu\nu},\chi_{(\mathrm{mat})}\right] \,.
\end{equation}
Here the Ricci scalar $\mathcal{R}$ can be taken either in the metric formalism as $\mathcal{R}= R \equiv g^{\mu \nu} R_{\mu\nu}$ formed of the Ricci tensor $R_{\mu\nu}$ which is computed from the Levi-Civita connection $\Gamma^\lambda{}_{\mu\nu}$ of the metric $g_{\mu\nu}$, or in the Palatini formalism as $\mathcal{R}= \widehat{R} \equiv g^{\mu \nu} \widehat{R}_{\mu\nu}$ formed of the Ricci tensor $\widehat{R}_{\mu\nu}$ that is computed from a symmetric affine connection $\widehat{\Gamma}^\lambda{}_{\mu\nu}$ that is independent of the metric $g_{\mu\nu}$.
The three model functions $\mathcal{A}$, $\mathcal{B}$, and $\mathcal{V}$ describe the nonminimal coupling to gravity, kinetic self-coupling, and the potential of the scalar field and specify a concrete theory. The fourth possible function $\sigma$ describes the nonminimal coupling between the scalar field and matter. 

In writing the above action, we have tacitly assumed that the theory is originally set up (and the system of physical units fixed) in the Jordan frame defined by $\sigma = 0$. In this frame, the effective gravitational constant varies with the scalar field dynamics according to the nonminimal coupling function $\mathcal{A}>0$, which takes on the role of the reduced Planck mass (squared). Giving the scalar field $\Phi$ mass dimension $1$ as usual, along with the coordinates having dimension $-1$, the model functions $\mathcal{A}$, $\mathcal{B}$, $\mathcal{V}$ are of mass dimension 2, 0, and 4, respectively. Furthermore, for consistency, the spacetime metric as well as connection are dimensionless. As a result, the Lagrangian density has mass dimension 4 and the action is also dimensionless. 

In both formalisms, the matter fields $\chi_{(\mathrm{mat})}$ in the matter action $S_{(\mathrm{mat})}$ are assumed to couple to the (Jordan frame) metric $g_{\mu\nu}$ only, and thus their energy-momentum tensor
\begin{align}
    T_{\mu\nu}^{(\mathrm{mat})} = - \frac{2}{\sqrt{-g}}\frac{\delta S_{(\mathrm{mat})}}{\delta g^{\mu\nu}}
\end{align}
obeys the conservation law 
\begin{align}
    \nabla_\mu T^{\mu\nu}_{(\mathrm{mat})} = 0 \,
\end{align}
with respect to the Levi-Civita covariant derivative \cite{Koivisto:2005yk}. The matter particles therefore follow the geodesics of the metric $g_{\mu\nu}$.

In the metric formalism, the variation of the action \eqref{action:flanagan:1} with respect to the metric $g_{\mu\nu}$ gives the tensor field equations
\begin{eqnarray}
\mathcal{A}G_{\mu\nu} &+& \left(\mathcal{A}^{\prime\prime}+\frac{\mathcal{B}}{2}\right)
 g_{\mu\nu} g^{\rho\sigma}\partial_\rho\Phi\partial_\sigma\Phi - \left(\mathcal{A}^{\prime\prime} +\mathcal{B} \right)
 \partial_\mu\Phi\partial_\nu\Phi 
+ \mathcal{A}^\prime\left( g_{\mu\nu}\Box\Phi - \nabla_\mu\partial_\nu\Phi\right) +  g_{\mu\nu}\mathcal{V} 
= T_{\mu\nu}^{(\mathrm{mat})} \,. 
\label{mf:tensor:equation}
\end{eqnarray}
where the Einstein tensor $G_{\mu\nu} = R_{\mu\nu} - \frac{1}{2}g_{\mu\nu}R$, the covariant derivative $\nabla_\mu$, and d'Alembertian operator $\Box=g^{\mu\nu}\nabla_\mu \nabla_\nu$ are all computed with the aid of the Levi-Civita connection $\Gamma^\lambda{}_{\mu\nu}$. Primes denote derivative w.r.t.\ the scalar field, e.g.\ $\mathcal{A}^\prime= \tfrac{d\mathcal{A}(\Phi)}{d\Phi}$, $\mathcal{A}^{\prime\prime}= \tfrac{d^2\mathcal{A}(\Phi)}{d\Phi^2}$.
The variation with respect to the scalar field gives an equation
\begin{eqnarray}
\label{mf:scalar:field:equation}
R\mathcal{A}^\prime + \mathcal{B}^\prime g^{\mu\nu}\partial_\mu\Phi\partial_\nu\Phi + 2\mathcal{B}\, \Box\Phi - 2\mathcal{V}^\prime = 0  \,
\end{eqnarray}
which depends on the Levi-Civita Ricci scalar $R$. We can ``debraid'' the equations by taking a contraction of Eq.\ \eqref{mf:tensor:equation},
\begin{equation}
\label{mf:contraction:tensor:equation}
-\mathcal{A}R + \mathcal{B}g^{\mu\nu} \partial_\mu \Phi \partial_\nu \Phi 
+ 3 \mathcal{A}^{\prime\prime}g^{\mu\nu} \partial_\mu \Phi \partial_\nu \Phi + 3\mathcal{A}^\prime \, \Box \Phi 
+ 4  \mathcal{V} = T^{(\mathrm{mat})} \,,
\end{equation}
and inserting $R$ into to the equation \eqref{mf:scalar:field:equation} to obtain the scalar field equation as
\begin{align}
\label{mf:scalar:field:equation:without:R}
\frac{2\mathcal{A}\mathcal{B} \negmedspace + \negmedspace 3\left( \mathcal{A}^\prime\right)^2}{\mathcal{A}}\, \Box\Phi 
+ \frac{\left( 2\mathcal{A}\mathcal{B} \negmedspace + \negmedspace 3\left( \mathcal{A}^\prime \right)^2 \right)^\prime}
{2\mathcal{A}} g^{\mu\nu} \partial_\mu\Phi \partial_\nu\Phi - \frac{2\left( \mathcal{A}\mathcal{V}^\prime \negmedspace 
- \negmedspace 2 \mathcal{A}^\prime \mathcal{V} \right)}{\mathcal{A}} &=  \frac{\mathcal{A}^\prime }{\mathcal{A}} T^{(\mathrm{mat})}  \,.
\end{align}
Now the trace of the matter energy-momentum appears as a source for the scalar field, on the RHS of \eqref{mf:scalar:field:equation:without:R}.

In the Palatini formalism, we begin by varying the action \eqref{action:flanagan:1} with respect to the metric. This yields
\begin{eqnarray}
\label{pf:tensor:equation}
 \mathcal{A} \widehat{G}_{\mu\nu}
 + \frac{1}{2} \mathcal{B} g_{\mu\nu} g^{\rho \sigma}\partial^{\rho} \Phi \partial_{\rho} \Phi 
 - \mathcal{B} \partial_{\mu} \Phi \partial_{\nu} \Phi +  g_{\mu\nu} \mathcal{V}
 = T_{\mu\nu}^{(\mathrm{mat})} \,,
\end{eqnarray}
where the Einstein tensor $\widehat{G}_{\mu\nu} = \widehat{R}_{\mu\nu}  - \frac{1}{2} g_{\mu\nu} \widehat{R} $ is assembled from the independent connection $\widehat{\Gamma}^{\lambda}{}_{\mu \nu}$, while indices are still raised, lowered, and contracted by the metric $g_{\mu\nu}$. Since the connection  $\widehat{\Gamma}^{\lambda}{}_{\mu \nu}$ is independent, we should vary the action \eqref{action:flanagan:1}  with respect to it as well, leading to the constraint 
\begin{equation}
\label{pf:connection:variation}
\widehat{\nabla}_{\sigma} \left(\sqrt{-g} \, \mathcal{A}  \, g^{\lambda \nu}\right) = 0\,.
\end{equation}
Here the covariant derivative $\widehat{\nabla}_{\rho}$ is taken with respect to the independent affine connection $\widehat{\Gamma}^{\lambda}{}_{\mu \nu}$. As long as the independent connection is assumed to be symmetric, the condition \eqref{pf:connection:variation} can be interpreted as a statement that the connection $\widehat{\Gamma}^{\lambda}{}_{\mu \nu}$ is a Levi-Civita connection for a conformally rescaled metric $\hat{g}_{\mu\nu}$, which is related to original metric $g_{\mu\nu}$ as
\begin{equation}
\label{eq:g hat}
\hat{g}_{\mu\nu} = (\mathcal{A}/M^2) \, {g}_{\mu\nu} \,,
\end{equation} 
where $M$ is a constant of mass dimension one intended to preserve dimensional consistency.
Since the two metrics $\hat{g}_{\mu\nu}$ and $g_{\mu\nu}$ are conformally related, so are the two respective connections obtainable from each other by a conformal transformation rule \begin{equation}
\label{hat:Gamma:Gamma}
\widehat{\Gamma}^{\lambda}{}_{\mu \nu} =\Gamma^{\lambda}{}_{\mu \nu} + \frac{\mathcal{A}^{\prime}}{2 \mathcal{A}}
\left(\delta_\mu^\lambda \, \partial_\nu \Phi + \delta_\nu^\lambda \, \partial_\mu \Phi - g_{\mu \nu} \, g^{\lambda \sigma} 
\, \partial_\sigma \Phi\right) \,.
\end{equation}
Essentially, the connection equation \eqref{pf:connection:variation} completely fixes the independent connection $\widehat{\Gamma}^{\lambda}{}_{\mu \nu}$, allowing it to be expressed in terms of the scalar field $\Phi$ and the original metric $g_{\mu\nu}$ via its Levi-Civita connection $\Gamma^{\lambda}{}_{\mu \nu}$ in \eqref{hat:Gamma:Gamma}. Thus, we can eliminate $\widehat{\Gamma}^{\lambda}{}_{\mu \nu}$ from the tensor field equation \eqref{pf:tensor:equation} by replacing $\widehat{G}_{\mu\nu}$ with the Einstein tensor $G_{\mu\nu}$ calculated from the Levi-Civita connection $\Gamma^{\lambda}_{\ \mu \nu}$ of the original metric $g_{\mu\nu}$, using another conformal relation \cite{Koivisto:2005yk}
\begin{eqnarray}
\label{pf:einstein:tensor:transformation}
\widehat{G}_{\mu\nu} &=& G_{\mu\nu}
+ \left( \frac{\mathcal{A}^{\prime\prime}}{\mathcal{A}} -\frac{3\left( \mathcal{A}^\prime \right)^2}{4\mathcal{A}^2} \right) g_{\mu\nu} g^{\rho\sigma}\partial_\rho\Phi \partial_\sigma\Phi
- \left( \frac{\mathcal{A}^{\prime\prime}}{\mathcal{A}} - \frac{3\left( \mathcal{A}^\prime \right)^2}{2\mathcal{A}^2}  \right) \partial_\mu\Phi \partial_\nu\Phi 
+ \frac{\mathcal{A}^\prime}{\mathcal{A}}g_{\mu\nu}\Box\Phi 
- \frac{\mathcal{A}^\prime}{\mathcal{A}}\nabla_\mu \partial_\nu\Phi 
\,.
\end{eqnarray}
In the end, the tensor field equation \eqref{pf:tensor:equation} that governs the dynamics of the metric $g_{\mu\nu}$ takes the form 
\begin{align}
\mathcal{A} G_{\mu\nu} + \left( \mathcal{A}^{\prime\prime} + \frac{\mathcal{B}}{2} - \frac{3 (\mathcal{A}^{\prime})^2}{4\mathcal{A}} \right)
 g_{\mu\nu} g^{\rho\sigma}\partial_\rho\Phi \, \partial_\sigma\Phi - \left( \mathcal{A}^{\prime\prime} +\mathcal{B} -\frac{3 (\mathcal{A}^{\prime})^2}{2\mathcal{A}} \right)
 \partial_\mu\Phi \, \partial_\nu\Phi \qquad
 \nonumber \\
+ \mathcal{A}^\prime\left( g_{\mu\nu}\Box\Phi - \nabla_\mu\partial_\nu\Phi\right) +  g_{\mu\nu}\mathcal{V} 
&=  T_{\mu\nu}^{(\mathrm{mat})} \,.
\label{pf:metric:equation}
\end{align}

Variation of the action \eqref{action:flanagan:1}  with respect to the scalar field gives
\begin{eqnarray}
\label{pf:scalar:field:equation}
\mathcal{A}^\prime \widehat{R}  + \mathcal{B}^\prime g^{\mu\nu}\partial_\mu\Phi\partial_\nu\Phi + 2\mathcal{B}\, \Box\Phi 
- 2\mathcal{V}^\prime = 0  \,,
\end{eqnarray}
which depends on the curvature scalar computed from the connection $\widehat{\Gamma}^{\lambda}{}_{\mu \nu}$. Again, to ``debraid'' the equations, we can contract Eq. \eqref{pf:tensor:equation} with the metric $g_{\mu\nu}$ to take the trace, and insert the result into Eq.
\eqref{mf:scalar:field:equation}, to obtain the scalar field equation
\begin{align}
\label{pf:scalar:field:equation:without:R}
2 \mathcal{B}  \, \Box \Phi + \left({\mathcal{B'}} + \frac{\mathcal{A'}}{\mathcal{A}} \mathcal{B} \right) 
g^{\rho\sigma}\partial_{\rho}\Phi \, \partial_{\sigma} \Phi  
- \frac{2 \left(\mathcal{A} \mathcal{V'} - 2 \mathcal{A'} \mathcal{V} \right)}{\mathcal{A}}
 &= \frac{\mathcal{A'}}{\mathcal{A}} T^{(\mathrm{mat})}\,.
\end{align}
Like in the metric formalism, this move introduces the trace of the matter energy-momentum as a source for the scalar field.

With a little inspection, it becomes apparent that the scalar-tensor field equations in the metric and Palatini formalisms are congruent to each other, and it is easy to incorporate both cases into unified expressions with the introduction of appropriate notation. We can write the tensor field equations in the metric  \eqref{mf:tensor:equation} and Palatini \eqref{pf:metric:equation} together as
\begin{align}
\mathcal{A} G_{\mu\nu} + \left( \frac{1}{2}\mathcal{B}+\mathcal{A}^{\prime\prime} - \delta_P \frac{3 (\mathcal{A}^{\prime})^2}{4\mathcal{A}} \right)
 g_{\mu\nu} g^{\rho\sigma}\partial_\rho\Phi \, \partial_\sigma\Phi - \left( \mathcal{B} + \mathcal{A}^{\prime\prime} -\delta_P \frac{3 (\mathcal{A}^{\prime})^2}{2\mathcal{A}} \right)
 \partial_\mu\Phi \, \partial_\nu\Phi \qquad & \nonumber \\
+ \mathcal{A}^\prime\left( g_{\mu\nu}\Box\Phi - \nabla_\mu\partial_\nu\Phi\right) +  g_{\mu\nu}\mathcal{V} 
&= T_{\mu\nu}^{(\mathrm{mat})} \,,
\label{mfpf:tensor:equation}
\end{align}
and the respective scalar field equations \eqref{mf:scalar:field:equation:without:R} and \eqref{pf:scalar:field:equation:without:R} as
\begin{align}
\label{mfpf:scalar:field:equation:without:R}
\frac{2\mathcal{A}\mathcal{B} + 3 \, \delta_m \left( \mathcal{A}^\prime\right)^2}{\mathcal{A}}\, \Box\Phi 
+ \frac{\left( 2\mathcal{A}\mathcal{B} + 3 \, \delta_m \left( \mathcal{A}^\prime \right)^2 \right)^\prime}
{2\mathcal{A}} g^{\rho\sigma} \partial_\rho\Phi \, \partial_\sigma\Phi - \frac{2\left( \mathcal{A}\mathcal{V}^\prime  
-  2 \mathcal{A}^\prime \mathcal{V} \right)}{\mathcal{A}} &= \frac{\mathcal{A'}}{\mathcal{A}}  T^{(\mathrm{mat})}\,.
\end{align}
Here the symbols
\begin{align}
\label{eq: metric Palatini delta}
    {\delta}_m &= \left\lbrace \begin{tabular}{ll}
        1 \,, & \textrm{metric} \\
        0 \,, & \textrm{Palatini}
    \end{tabular} \right. 
    \,, \qquad
    {\delta}_P = \left\lbrace \begin{tabular}{ll}
        0 \,, & \textrm{metric} \\
        1 \,, & \textrm{Palatini}
    \end{tabular} \right. 
\end{align}
switch on the two extra terms that appear in the tensor field equation in the Palatini case and the two extra terms that appear in the scalar field equation in the metric case. These equations generalize the Brans-Dicke-type case, where the change in the formalism is reduced to a simple shift in the Brans-Dicke parameter $\omega$ \cite{Lindstrom:1975ry,Lindstrom:1976pq,vandenBergh:1981,Burton:1997pe}. Here we see that for general model functions, the difference in metric and Palatini is slightly more involved, and can be expressed as
\begin{align}
    \Bcal_{\rm Palatini} &= \Bcal_{\rm metric} + \frac{3 (\Acal)'}{2 \Acal} \,.
\end{align}
Thus, at least classically, the Palatini formalism does not introduce a new class of theories compared to the metric formalism, but simply corresponds to a reshuffle of the degrees of freedom provided by the choice of model functions.\footnote{A similar metric-Palatini relationship can be generalized to more elaborate Horndeski-type of theories as well \cite{Helpin:2019kcq}.} However, we do not know {\it a priori} which formalism is more physically motivated, and so when motivating or postulating a model, e.g.\ coupling the Standard Model Higgs field nonminimally to curvature, the distinction is relevant \cite{Bauer:2008zj}. In what follows, we keep the delta symbols \eqref{eq: metric Palatini delta} explicit in order to make the formulae maximally transparent and to facilitate the comparison between the metric and Palatini cases.

We assume that the effective gravitational constant is always positive and the potential is nonnegative, i.e.
\begin{align}
\label{eq: assumptions on the model functions}
    \mathcal{A}>0 \,, \qquad \mathcal{V}\geq 0 \,.
\end{align}
This is to avoid any instability scenarios. The condition that the scalar field is not a ghost can be read off from the factor multiplying the $\Box\Phi$ term in the scalar field equation \eqref{mfpf:scalar:field:equation:without:R}, 
\begin{equation} 
\label{eq: no ghost condition}
2\mathcal{A}\mathcal{B} + 3 \, \delta_m \left( \mathcal{A}^\prime\right)^2\,  > 0 \,.
\end{equation}
We see that in the Palatini case, the requirement boils down to the correct sign of the kinetic term in the action, while in the metric case, the condition appears more involved. Naively, the condition gives the correct signs that provide stability, a generalization of $\ddot{\Phi} \sim - \mathcal{V}^\prime$ in the minimally coupled case. However, there is also an intuitive, unified way to understand this condition in both the Palatini and the metric cases: the non-ghost requirement corresponds to a positive-definite field space metric \cite{Kuusk:2015dda,Hohmann:2016yfd}.

\section{FLRW cosmology and slow roll}
\label{sec:FLRWcosmology}

Let us now narrow our focus upon cosmological dynamics and consider a spatially homogeneous and isotropic flat ($k=0$) spacetime, given by the  Friedmann-Lema\^itre-Robertson-Walker line element 
\begin{equation}
\label{FLRW}
\mathrm{d}s^2 = -\mathrm{d}t^2 + a(t)^2 \mathrm{d} \mathbf{x}^2 \,,
\end{equation}
where $t$ corresponds to the cosmic time.
The homogeneity and isotropy assumptions demand that the scalar field only depends on the cosmological time, $\Phi(t)$. In the context of inflation, the other fields can be neglected, as their densities are quickly diluted.

Substituting these assumptions into the tensor and scalar field equations \eqref{mfpf:tensor:equation} and \eqref{mfpf:scalar:field:equation:without:R} yields the following set of equations to model inflation:
\begin{subequations}
\label{eq:FLRW equations}
\begin{align}
\label{mpf:first:friedmann:1} 
H^2 &= -\frac{\mathcal{A}^\prime}{ \mathcal{A} } H \dot{\Phi} + \frac{ 2\mathcal{A} \mathcal{B} - 3 \delta_P (\mathcal{A}')^2}{ 12\mathcal{A}^2 } \dot{\Phi}^2 
+ \frac{\mathcal{V}}{3\mathcal{A}}  \,, \\
\label{mpf:second:friedmann:2} 
2 \dot{H}  +  3H^2 &= -2\frac{\mathcal{A}^\prime}{\mathcal{A}} H \dot{\Phi} 
- \frac{4 \mathcal{A} \mathcal{A}'' + 2 \mathcal{A} \mathcal{B} - 3 \delta_P (\mathcal{A}')^2 }{4\mathcal{A}^2} \dot{\Phi}^2 
- \frac{\mathcal{A}^\prime}{\mathcal{A}}\ddot{\Phi}
+ \frac{\mathcal{V}}{\mathcal{A}}   \,, \\
\label{mpf:scalar:field:equation:friedmann:1}
\ddot{\Phi} &= -3 H\dot{\Phi} - \frac{\mathcal{A} \mathcal{B}' + \mathcal{A}' \mathcal{B} 
+ 3 \, \delta_m \mathcal{A}^\prime \mathcal{A}'' }{2 \mathcal{A} \mathcal{B} 
+ 3 \, \delta_P (\mathcal{A}^\prime)^2}\dot{\Phi}^2  
- \frac{ 2 \mathcal{A} \mathcal{V}^\prime  - 4 \mathcal{V} \mathcal{A}^\prime }{2 \mathcal{A} \mathcal{B} 
+ 3 \, \delta_m (\mathcal{A}^\prime)^2} \,,
\end{align} 
\end{subequations}
where dots denote derivatives with respect to the time variable and the expansion is characterized by the Hubble function $H=\tfrac{\dot{a}}{a}$.
Essentially, the system is 2-dimensional, since we can algebraically express $H$ from the Friedmann constraint \eqref{mpf:first:friedmann:1} and substitute into \eqref{mpf:scalar:field:equation:friedmann:1} which gives a 2nd order differential equation for the scalar field only. The other Friedmann equation \eqref{mpf:second:friedmann:2} derives from \eqref{mpf:first:friedmann:1} and \eqref{mpf:scalar:field:equation:friedmann:1}. Thus once the dynamics of $\Phi$ is known, the evolution of $H$ is determined by the constraint \eqref{mpf:first:friedmann:1}.
Following Ref.\ \cite{Jarv:2021ehj} we can interpret the first two terms on the RHS of \eqref{mpf:scalar:field:equation:friedmann:1} as friction, while the third plays the role of the gradient of the effective potential (like a ``force'' divided by the effective ``mass''),
\begin{align}
\label{eq: Veff}
    \Veff &= \frac{\mathcal{V}}{\mathcal{A}^2} \,,\\
\label{eq: m_eff}
    \meff &= \frac{2 \mathcal{A}\mathcal{B} + 3 \delta_m (\mathcal{A'})^2}{2 \mathcal{A}^3} \,.
\end{align}
These two quantities rule the scalar field dynamics. The fixed points of the scalar field dynamics, i.e.\ where the scalar field evolution stops, occur at the extrema of the effective potential or at the values where the effective mass diverges, more precisely
\begin{align}
\label{eq:general_fixed_point_condition}
    \frac{\Veff'}{\meff} \Big|_{FP} &= \frac{ 2 \mathcal{A} \mathcal{V}^\prime  - 4 \mathcal{V} \mathcal{A}^\prime }{2 \mathcal{A} \mathcal{B} + 3 \, \delta_m (\mathcal{A}^\prime)^2} \Big|_{FP} = 0 \,.
\end{align}
Here the effective mass describes the scalar field cosmological dynamics and is not directly related to the mass of the scalar quantum particle. Rather the effective mass is proportional to the one-dimensional analog of the field space metric, which for a single field is reduced to a scalar function \cite{Kuusk:2015dda,Hohmann:2016yfd}. For example, we can understand that given a ``force'' the scalar field evolution can slow down because the field gets more ``massive'' or equivalently because the invariant ``distance'' in the field space increases. 
Note that while the effective potentials coincide in the metric and Palatini formalism, the effective mass is different. 
However, this difference amounts to a positive factor only, since all terms in \eqref{eq: m_eff} are positive by definition. In particular, the metric and Palatini cases possess the same fixed points with the same properties. This feature will be important later.

The rate of expansion can be conveniently measured in terms of the effective barotropic index
\begin{align}
\label{eq:weff}
    \weff &= -1 - \frac{2 \dot{H}}{3 H^2} = -1 - \frac{2 \left[2 \mathcal{A} \mathcal{A}^{\prime} \ddot{\Phi} - 2 H \mathcal{A} \mathcal{A}^{\prime} \dot{\Phi} + \left( 2 \mathcal{A} \mathcal{B} + 2 \mathcal{A} \mathcal{A}^{\prime\prime} - 3 \delta_{P} \left(\mathcal{A}^{\prime}\right)^{2}\right)\dot{\Phi}^{2}\right]}{12 H \mathcal{A} \mathcal{A}^{\prime} \dot{\Phi} - 4 \mathcal{A} \mathcal{V} - \left(2 \mathcal{A} \mathcal{B} - 3 \delta_{P} \left(\mathcal{A}^{\prime}\right)^{2}\right) \dot{\Phi}^{2} } \,.
\end{align}
In the case of minimal coupling ($\mathcal{A}\equiv 1$), $\weff$ varies between $-1$ (achieved in the potential dominated regime where the scalar field derivatives can be neglected over the potential), and $+1$ (achieved in the kinetic regime where the potential can be neglected over the derivatives). In the case of nonminimal coupling, the effective barotropic index can take a wider range of values, including the superaccelerating regime where $\weff<-1$ and $\dot{H}>0$, as well as superstiff regime where $\weff>1$. When the value of $\weff$ is constant, we can integrate the left side of \eqref{eq:weff} to get 
\begin{subequations}    
\label{eq:weff_to_a(t)}
\begin{align}
    H &= \frac{2}{3(1+\weff)(t-t_0)} \,, & a &= a_0 (t-t_0)^{\frac{2}{3(1+\weff)}} \,, & \weff &\neq -1 \,,  \\
    H &= H_0 \,, & a &=a_0 \, e^{H_0 t} \,, & \weff &= -1 \,, 
\end{align}
\end{subequations}
where $t_0$ and $a_0$ are integration constants. Knowing $H(t)$ then allows us to integrate Eq.\ \eqref{mpf:first:friedmann:1} and find $\Phi(t)$. The scalar field finite fixed points are always de Sitter with $\weff=-1$ \cite{Burd:1991ns,Gunzig:2000kk,Gunzig:2000ce,Jarv:2008eb,Jarv:2010zc,Jarv:2014hma}, but as we will see later, the scalar field asymptotic states can correspond to power law expansion with different values of $\weff$.

To resolve the horizon problem and to generate a nearly scale-invariant spectrum of perturbations that made an imprint on the observations today, the scalar field has to usher a nearly de Sitter expansion during inflation. This is approximated by the slow roll regime where the time derivatives of the field can be considered to be small. Dropping the $\dot{\Phi}$ term in the Friedmann equation~\eqref{mpf:first:friedmann:1} and analogously neglecting the $\ddot{\Phi}$ and $\dot{\Phi}^2$ terms in the scalar field equation \eqref{mpf:scalar:field:equation:friedmann:1} yields the following slow roll conditions:
\begin{subequations}
\label{eq:slowroll}
    \begin{align}
    \label{eq:slowroll1}
    H^2 &= \frac{ \mathcal{V}}{3  \mathcal{A}} \,, \\
    \label{eq:slowroll2}
    3 H\dot{\Phi} &= \frac{ 2\mathcal{A} \mathcal{V}^\prime  - 4 \mathcal{V} \mathcal{A}^\prime }{2 \mathcal{A} \mathcal{B}
    +  3 \delta_m (\mathcal{A}^\prime)^2}  \,,
\end{align}
\end{subequations}
We kept the $\dot{\Phi}$ term in the second equation \eqref{eq:slowroll2}, since otherwise dropping it would have simply reduced the expression to the de Sitter fixed point condition \eqref{eq:general_fixed_point_condition}. Therefore \eqref{eq:slowroll2} characterizes slow roll as a small deviation from the exact de Sitter behavior. 
Note that the slow roll expression is slightly different in the metric and Palatini formalism. In the metric formalism it reproduces the earlier result and matches the Einstein frame slow roll conditions translated into Jordan frame \cite{Chiba:2008ia,Akin:2020mcr,Jarv:2021qpp,Karciauskas:2022jzd}.\footnote{Some authors have proposed ``generalized slow roll conditions'' in the Jordan frame but that approach has a flaw, see Ref.\ \cite{Jarv:2021qpp}.} Up to our knowledge, the Jordan frame slow roll conditions in the Palatini formulation have not been written down before.

As we assume that $\mathcal{A}>0$ and $\mathcal{V}\geq0$, the Friedmann constraint \eqref{mpf:first:friedmann:1} puts a bound on the allowed values of the cosmological variables,
\begin{align}
\label{eq:Friedmann_bound}
    1 + \frac{\mathcal{A}^\prime \dot{\Phi} }{ \mathcal{A} H } - \frac{ \left(2\mathcal{A} \mathcal{B} - 3 \delta_P (\mathcal{A}')^2\right) \dot{\Phi}^2 }{ 12\mathcal{A}^2 H^2 } \geq 0 \,,
\end{align}
i.e.\ for a given values of $\Phi, \dot{\Phi}$ the range of possible expansion rates $H$ is limited. This relation establishes a boundary in the phase space of the dynamics \cite{Faraoni:2005vc,Jarv:2009zf}. The equal sign in \eqref{eq:Friedmann_bound} corresponds to an utmost kinetic regime for the scalar field, where the role of the potential can be completely neglected in the Friedmann equation \eqref{mpf:first:friedmann:1}.

\section{Inflationary observables}
\label{sec:observables}

The key success of the inflationary paradigm lies in its predictive power with respect to the perturbations which can be observed in the sky today. A very useful tool to derive the spectrum of perturbations for any scalar-tensor model is provided by the concept of invariants, applicable both in the metric \cite{Jarv:2016sow} as well as Palatini \cite{Jarv:2020qqm} formalisms. In order to proceed, let us briefly review of how and why this approach works.

As is well known, the structure of the action \eqref{action:flanagan:1} is preserved under a \emph{frame transformation} \cite{Flanagan:2004bz,Jarv:2014hma,Jarv:2015kga,Burns:2016ric,Kozak:2018vlp}:
\begin{subequations}
\label{eq:transformations}
\begin{align}
    g_{\mu\nu} &= e^{2 \bar{\gamma}(\bar{\Phi})} \bar{g}_{\mu\nu} \,,  \label{subeq1}
 \\
    \Phi &= \bar{f}(\bar{\Phi}) \,. \label{subeq2}
\end{align}
\end{subequations}
Breaking up the frame transformation in its constituent parts, we see that it is made up of a conformal transformation\footnote{More appropriately \emph{Weyl rescaling}, although the term ``conformal transformation" is usually employed in the literature to refer to a field-dependent redefinition of the metric that maintains its signature.} \eqref{subeq1} and a field redefinition \eqref{subeq2}. This transformation has no explicit dependence on spacetime, and so only the model functions $\mathcal{A}$, $\mathcal{B}$, $\mathcal{V}$, and $\sigma$ are being transformed \cite{Jarv:2014hma,Jarv:2016sow, Jarv:2020qqm} (not the coordinates). As a result, there are no degrees of freedom introduced by the transformation, and so its effects can be captured by a set of transformation rules for curvature invariants. Nonetheless, the form of the action is nonetheless unchanged: the action still does not feature higher derivatives.

Thankfully, there is a set of simple quantities that remain invariant under the transformations above \cite{Jarv:2014hma,Jarv:2016sow,Jarv:2020qqm}, given by
\begin{align}
	\label{I_m}
    \IM(\Phi) &= \frac{\mathrm{e}^{2\sigma(\Phi)}}{\mathcal{A}(\Phi)/ {M^2}},\\
	\label{I_V}    
    \IV(\Phi) &= \frac{\mathcal{V}(\Phi)}{\left(\mathcal{A}(\Phi)\right)^2},\\
	\label{I_Phi} 
	\IP(\Phi) &=  \int\!\mathrm{d}\Phi\,\sqrt{\frac{\mathcal{B}(\Phi)}{\mathcal{A}(\Phi)} 
		+  \frac{3}{2} \, \delta_{m}  \, \left(\frac{\mathcal{A}'(\Phi)}{\mathcal{A}(\Phi)}\right)^2} \,,
\end{align} 
meaning the numerical value of these quantities at a spacetime point does not change, $\bar{\mathcal{I}}_i(\bar{\Phi})={\mathcal{I}_i}({\Phi})$. 
Therefore, by definition, these quantities have zero scaling dimension, as they do not pick up any conformal factors $e^{2\bar \gamma(\bar \Phi)}$ under a conformal transformation. They also have zero mass dimension: this is not necessarily true by construction, so in order to achieve a concordance between the two notions of dimension, we have scaled them with an arbitrary dimensionful parameter $M$.
 
In both the metric and Palatini formalisms, the quantity $\IP$ provides an invariant description of the scalar degree of freedom in the theory \cite{Jarv:2014hma,Kozak:2018vlp}. Having an identically constant $\IP$ corresponds to a nondynamical scalar field, indicating that the theory is exactly equivalent to general relativity plus a cosmological constant. Negative values under the square root in \eqref{I_Phi} signal that the scalar is a ghost, cf.\ \eqref{eq: no ghost condition}. In essence, the invariant potential $\IV(\IP)$ combines the effective potential \eqref{eq: Veff} with the information of the effective mass \eqref{eq: m_eff}. When the effective mass is larger and thus the field $\Phi$ evolves ``slower'', the invariant potential becomes more ``flat'' since the invariant field $\IP$ gets 
``stretched out'' compared to the original field $\Phi$.

We may analogously introduce an invariant metric $\hat{g}_{\mu\nu}=\mathcal{A}g_{\mu\nu}$ which coincides with \eqref{eq:g hat} and is similarly unaffected by the transformation \eqref{eq:transformations} \cite{Jarv:2014hma,Kozak:2018vlp}. Using the invariant metric and the invariant quantities it is possible to rewrite the equations of inflationary cosmology of all models described by \eqref{action:flanagan:1} in a unified compact way. In order to avoid a nontrivial lapse function in $\hat{g}_{\mu\nu}$ and to maintain the FLRW form of \eqref{FLRW}, i.e.
\begin{equation}
\label{FLRW:hat}
\mathrm{d}\hat{s}{}^2 = - \mathrm{d}\hat{t}{}^2 + \hat{a}(\hat{t})^2 \mathrm{d} \mathbf{x}^2 \,,
\end{equation}
it is helpful to use a rescaled time coordinate, $\hat{t}=(\sqrt{\mathcal{A}}/{M}) t$ and scale factor $\hat{a}=(\sqrt{\mathcal{A}}/{M}) a$.
Then the \emph{invariant} Hubble parameter $\widehat{H}$ calculated in terms of the invariant variables is related to the Hubble parameter $H$ calculated in the frame defined by $g_{\mu\nu}$ as 
\begin{align}
\label{eq: H:hat}
\widehat{H} \equiv \frac{1}{\sqrt{\mathcal{A}}}\left(H + \frac{1}{2}\frac{\mathcal{A}^\prime}{\mathcal{A}}\dot{\Phi}\right) \,.
\end{align}
This definition is akin to that of a \emph{frame-covariant time derivative} \cite{Postma:2014vaa, Karamitsos:2017elm}. Such a derivative takes into account the scaling dimension of the quantity it acts upon, and it respects field-dependent unit transformations (i.e.\ conformal transformations) to \emph{any} frame: \eqref{eq: H:hat} is the particular case for the transformation from the Jordan to the Einstein frame. 

Substituting the invariant form of FLRW metric \eqref{FLRW} into equations \eqref{mpf:first:friedmann:1}--\eqref{mpf:scalar:field:equation:friedmann:1} yields
\begin{subequations}
\label{eq:FLRW equations with invariants}
\begin{align}
\label{Friedmanns_constraint_in_terms_of_invariants} 
\widehat{H}^2 =& \frac{1}{3}\left(\frac{\DD \IP }{\DD\hat{t}} \right)^2 + \frac{\IV}{3} \,, \\
\label{Friedmanns_second_equation_in_terms_of_invariants}
2\frac{\DD \widehat{H}}{\DD\hat{t}}  + 3 \widehat{H}^2  =&  -\left(\frac{\DD\mathcal{I}_\Phi}{\DD\hat{t}}\right)^2 + \IV   \,, \\
\label{Phi_cosmo_equation_in_terms_of_invariants}
\frac{\DD^2 \mathcal{I}_\Phi}{\DD\hat{t}^2} =& -3\widehat{H} \frac{\DD\IP}{\DD\hat{t}} - \frac{1}{2}\frac{\DD \IV}{\DD\IP} \,.
\end{align}   
\end{subequations}
Thus all models, both metric and Palatini and with arbitrary coupling functions, follow the same equations where the dynamics of the invariant field $\mathcal{I}_\Phi$ is determined by the invariant potential $\mathcal{I}_{V}$, and the expansion $\widehat{H}$ follows algebraically from the evolution of the invariant field $\mathcal{I}_\Phi$. The third invariant quantity $\IM$ which describes matter coupling to gravity does not play a role in the inflationary epoch of the universe since matter gets diluted away by the near exponential expansion. 

The equations \eqref{eq:FLRW equations with invariants} coincide with the FLRW equations for a minimally coupled scalar field (or the Einstein frame of a nonminimally coupled scalar field). This affords us the ability to simply invoke the well known results from that case. 
Assuming the slow-roll conditions, it is possible to rewrite the potential slow-roll parameters as \cite{Kuusk:2016rso,Jarv:2016sow,Karam:2017zno}
\begin{align}
	\label{epsilon}
	\hat{\epsilon} &= \frac{1}{2}\left(\frac{\DD \ln \IV}{\DD \IP}\right)^2\,,\\
	\label{eta}
	\hat{\eta} &= \frac{1}{\IV}\frac{\DD^2 \IV}{\DD \IP^2}\,.
\end{align}
Then for the perturbations in the slow roll regime the scalar spectral index $n_s$, the tensor-to-scalar ratio $r$, and the amplitude of the scalar power spectrum $A_s$ follow in up to the first order in these parameters as \cite{Jarv:2016sow,Karam:2017zno}
\begin{align}
	\label{ns}
	n_s &= 
	1 -3 \left(\frac{\mathrm{d} \ln \IV}{\mathrm{d} \IP}\right)^2+2\,\frac{1}{\IV}\frac{\mathrm{d}^2 \IV}{\mathrm{d} \IP^2}  \,,
	\\
 \label{r}
	r &= 
	8\left(\frac{\mathrm{d} \ln \IV}{\mathrm{d} \IP}\right)^2 \,,
	\\
	\label{scalar.amplitude}
	A_s &= \frac{\IV}{12\pi^2}\left(\frac{\mathrm{d} \ln \IV}{\mathrm{d} \IP}\right)^{-2} \,.
\end{align}
Note that all of the above observables are calculated at field value $\mathcal{I}_\Phi^*$ corresponding to the moment where the perturbations exit the horizon. The number of e-foldings, characterizing the extent of expansion during the inflation from the start till the end at $\hat{\epsilon}=1$ ($\hat{w}_{\mathrm{eff}}=-\frac{1}{3}$), is given by
\begin{equation}
	\label{number.of.efolds}
	\widehat{N}{}^{\text{end}}_{*}=  \ln \hat{a}{}_{\text{end}} - \ln \hat{a}{}_{*} = \ln \left[ \frac{\hat{a}{}_{\text{end}}}{\hat{a}{}_{*}} \right] \approx \int\limits_{\IP^\text{end}}^{\IP^*}
	\IV(\IP)\left(\frac{\mathrm{d}\IV(\IP)}{\mathrm{d}\IP}\right)^{-1}\mathrm{d} \IP\,,
\end{equation}
where $\IP^\text{end}$ and $\IP^*$ (also $\hat{a}{}_{\text{end}}$ and $\hat{a}{}_{*}$) are the field values (scale factors) at the end and the beginning of inflation, respectively. The measure of inflation in terms of the original metric and original field is slightly different \cite{Burns:2016ric,Karam:2017zno,Racioppi:2021jai},
\begin{align}
\label{eq: N in Jordan and Einstein}
    {N}{}^{\text{end}}_{*}= \ln \left[ \frac{{a}{}_{\text{end}}}{{a}{}_{*}} \right] \approx \widehat{N}{}^{\text{end}}_{*} - \frac{1}{2} \ln \left[ \frac{\mathcal{A}(\Phi_{\text{end}})}{\mathcal{A}(\Phi_{*})} \right] \,.
\end{align}
However, in the slow roll approximation, the last term is usually considered to be subdominant and neglected. Thus the formulae above offer a quick algorithm to estimate the inflationary observables for any scalar-tensor model without repeating the perturbative calculations in every possible case separately \cite{Jarv:2016sow,Jarv:2020qqm}. The resulting observables can then be compared with the most recent best fit combined constraints (at $68 \%$ level) \cite{Planck:2018vyg} 
\begin{subequations}
\label{eq:obs1}
\begin{align}
\label{eq: Planck ns}
    n_s &= 0.9649 \pm 0.0042 \,, \\
\label{eq: Planck r}
    r &< 0.056 \,, \\
\label{eq: Planck A_s}
    A_s &= (2.1 \pm 0.059) \times 10^{-9} \, .
\end{align}
\end{subequations}

A deeper reason why the algorithm works for all scalar-tensor models is that the fixed points of the invariant field $\mathcal{I}_\Phi$ where its dynamics halts always entail de Sitter-type expansion \cite{Jarv:2014hma}. This is an invariant property of the system, and does not depend on whether we measure expansion with respect to the original metric or in conformally related metric and coordinates \eqref{FLRW:hat}. If the scalar field stops in time $t$, $\dot{\Phi}=0$, then $\IP$ also stops in time $\hat{t}$. Then \eqref{Friedmanns_constraint_in_terms_of_invariants} tells that $\widehat{H}{}$ is constant, which means that $H$ must also be constant due to Eq.\ \eqref{eq: H:hat}. Moreover, the stability properties of the scalar field fixed points are also invariant, since the eigenvalues of the linearized system near the fixed point can be given in terms of the invariant quantities \eqref{I_V}--\eqref{I_Phi} \cite{Jarv:2014hma}. Slow roll is by definition a nearly de Sitter regime, and thus there is no wonder that it also nearly invariant.

At this point, an attentive reader may raise a concern about what happens if the model functions harbor singularities for some values of the scalar field. In the case the map \eqref{I_Phi} from the original field to the invariant field has singularities which arise from the singularities in the original coupling functions ${\mathcal A}(\Phi)$ and ${\mathcal B}(\Phi)$, the invariant potential will not fully capture the entirety of the theory (since it is not possible to integrate over the singularities). Instead, the integral will be \emph{between} the singularities, meaning that the new field takes on the role of a \emph{chart} parametrizing a particular subdomain of the original space. Therefore, the new field is not enough to parametrize the whole field space: instead, multiple charts, each parametrized by a different invariant field, are necessary to span the whole of the field space \cite{Karamitsos:2019vor}. These different charts (once inverted and inserted into the original potential) correspond to different invariant potentials, and therefore different 
physical behaviors. However, because it is not possible to cross a singularity in the field space, these different 
domains have no way to ``communicate.' As such, choosing one particular domain in which the field is known to evolve, and therefore finding the associated invariant field, is enough to avoid any issues with singularities. One can also see that the stability properties of the fixed points are still retained even when the reparametrization of the scalar field is singular at that point, e.g.\ in the Brans-Dicke like parametrization $\omega\to \infty$ limit \cite{Jarv:2014laa,Jarv:2015kga}.

It is obvious from the discussion above, that metric or Palatini models that share invariant potentials $\IV(\IP)$ which are proportional to each other, will predict the same values for the inflationary observables like $n_s$ and $r$ \cite{Jarv:2016sow,Jarv:2020qqm}. This leads to a significant degeneracy in the zoo of models, as there are many models postulated with different actions that give identical inflationary predictions. The purpose of the present article is to try to unravel the degeneracy by looking beyond the slow roll. Given two models with proportional invariant potentials, what features in the overall dynamics of the system are the same, and what features can be different? A helpful tool to understand this issue is provided by the method of dynamical systems, as we may draw all solutions (corresponding to all possible initial conditions in the phase space) onto a single global phase portrait.

\section{Inflation as a dynamical system}
\label{sec:dynsys}

We will now present the study of inflation as a dynamical system, extending the approach of Refs.\ \cite{Dutta:2020uha,Jarv:2021qpp} to include the Palatini case for direct comparisons. First, for convenience and consistency, we will nondimensionalize the system by defining the  following dimensionless variables:
\begin{align}
\label{eq:dynamical variables}
\phi = \frac{\Phi}{M} \,, \qquad z=\frac{\dot{\phi}}{H} \, ,
\end{align}
where $M$ is a constant of mass dimension one. Since the variable $\phi$ is strictly proportional to the scalar field $\Phi$, this ensures that the dynamical system is closed for any combination of the model functions $\mathcal{A}(\Phi), \mathcal{B}(\Phi), \mathcal{V}(\Phi)$.

At this point, it is important to reflect on the meaning of the arbitrary mass parameter, which we introduced in order to keep dimensions consistent in the phase space. The reason we did not use the effective reduced Planck mass $\MP$ in its place is because the very notion of the Planck mass rests upon the model functions of the theory. While $\MP$ is of course a constant, its value can only be determined after choosing a frame and then examining the low-energy limit of the action. This does not change the physics, but since the definition of the Planck mass necessarily invites us to choose a frame, we replace it with an entirely arbitrary constant mass scale $M$ whose numerical value is not defined as the limit of some frame representation of the model under consideration. Nonetheless, depending on the particulars of the model, it may be convenient to define $M$ either as coinciding with the Planck mass for either the Jordan or the Einstein frame, or some other function of the Planck mass. In general, however, it is helpful to not assume it takes on any physically meaningful value, and still denote it by $M$.

It is useful to measure the expansion of the universe in e-folds $N=\ln a$, which are also dimensionless. This introduces the derivative of dimensionless ``time'' as
\begin{align}
\label{eq:dN and dt}
\frac{\DD}{\DD N} = \frac{1}{H}\frac{\DD}{\DD t} \, .
\end{align}
From this it follows that $\tfrac{\DD\phi}{\DD N}=z$, and so the number of expansion e-folds along a particular trajectory in the phase space can be calculated easily as 
\begin{equation}
\label{eq: N integral}
N = \int \DD N = \int H \DD t = \int \frac{H}{\dot{\phi}} \DD \phi = \int \frac{1}{z} \DD \phi \,.
\end{equation}
Here we assume an expanding universe, $\dot{H}>0$, which means that $N>0$ for both increasing and decreasing $\phi$, since in the latter case also $z<0$. Starting from the derivatives with Eq.\ \eqref{eq:dN and dt} and substituting in $\ddot{\phi}$, $\dot{H}$ and finally $H^2$ from Eqs.\ \eqref{eq:FLRW equations}, we arrive at the full form of the dynamical system:
\begin{subequations}
\label{eq:dynsys}
\begin{align}
    \frac{\DD \phi}{\DD N} =& z \,, 
    \label{eq:dynsys:phi}\\
    \frac{\DD z}{\DD N} =& \frac{1}{4 \mathcal{A}^2 \left(2 \mathcal{A} \mathcal{B} + 3 \delta_m (\mathcal{A}')^{2} \right)} 
     \Big[ 
     - 24 \mathcal{A}^{3} \left(\mathcal{A} \tfrac{\mathcal{V}^{\prime}}{\mathcal{V}} - 2 \mathcal{A}^{\prime} \right) 
     - 12 \mathcal{A}^{2} z \left(3 \mathcal{A} \mathcal{A}^{\prime} \tfrac{\mathcal{V}^{\prime}}{\mathcal{V}} + 2 \mathcal{A} \mathcal{B} - 6 \left(\mathcal{A}^{\prime}\right)^{2} \right) 
     \nonumber \\
     & \qquad \qquad \qquad \qquad \qquad \qquad 
     + 4 \mathcal{A} z^{2} \left(\mathcal{A}^{2} \mathcal{B} \tfrac{\mathcal{V}^{\prime}}{\mathcal{V}} - \mathcal{A}^{2} \mathcal{B}^{\prime} - 3 \mathcal{A} \left(\mathcal{A}^{\prime}\right)^{2} \tfrac{\mathcal{V}^{\prime}}{\mathcal{V}} - 7 \mathcal{A} \mathcal{A}^{\prime} \mathcal{B} + 6 \left(\mathcal{A}^{\prime}\right)^{3} \right)
     \nonumber \\
     & \qquad \qquad \qquad \qquad \qquad \qquad
     + 2 \mathcal{A} z^{3} \left(2 \mathcal{A} \mathcal{A}^{\prime\prime} \mathcal{B} + \mathcal{A} \mathcal{A}^{\prime} \mathcal{B} \tfrac{\mathcal{V}^{\prime}}{\mathcal{V}} - \mathcal{A} \mathcal{A}^{\prime} \mathcal{B}^{\prime} + 2 \mathcal{A} \mathcal{B}^{2}  - 3 \left(\mathcal{A}^{\prime}\right)^{2} \mathcal{B} \right) 
     \nonumber \\
     & \qquad \qquad \qquad \qquad \qquad \qquad
     + 6 \delta_{m} \mathcal{A} \mathcal{A}^{\prime} z \left(- 6 \mathcal{A} \mathcal{A}^{\prime} - 2 \mathcal{A} \mathcal{A}^{\prime\prime} z - 4 \left(\mathcal{A}^{\prime}\right)^{2} z + \mathcal{A}^{\prime} \mathcal{B} z^{2}\right) 
     \nonumber \\
     & \qquad \qquad \qquad \qquad \qquad \qquad
      - 3 \delta_{P} \left(\mathcal{A}^{\prime}\right)^{2} z^{2} \left(2 \mathcal{A}^{2} \tfrac{\mathcal{V}^{\prime}}{\mathcal{V}} - 4 \mathcal{A} \mathcal{A}^{\prime} + \mathcal{A} \mathcal{A}^{\prime} \tfrac{\mathcal{V}^{\prime}}{\mathcal{V}} z + 2 \mathcal{A} \mathcal{B} z - 2 \left(\mathcal{A}^{\prime}\right)^{2} z\right)
  \Big] \,.
  \label{eq:dynsys:z}
\end{align}
\end{subequations}
The system is two dimensional and fully encapsulates the dynamics of Eqs. \eqref{eq:FLRW equations} in terms of the new variables. The overall factor of the second equation \eqref{eq:dynsys:z} is always positive due to the no ghost condition \eqref{eq: no ghost condition}. 

We should note that the system is singular at $\tfrac{\mathcal{V}^{\prime}}{\mathcal{V}}=0$, e.g.\ at the minimum $\Phi=0$ of the power law potentials $\mathcal{V}\sim \Phi^n$ with $n>1$. However, this minimum is reached only in the aftermath of inflation, at which point the dynamical system is no longer an accurate description of the physics anyway. This is because reheating and other thermal processes become important when the scalar field starts to oscillate around the minimum of the potential and the matter sector is no longer diluted. The advantage of the current variables lies in the scalar field asymptotes: the infinite value of the potential on the asymptotes does not make the system singular, but instead may feature a fixed point from where it is possible to trace the inflationary heteroclinic orbit \cite{Jarv:2021qpp}.  

In terms of the chosen variables \eqref{eq:dynamical variables}, the barotropic index \eqref{eq:weff} is expressed as follows:
\begin{align}
    \weff =& -1 - \frac{1}{6 \mathcal{A}^2 \left(2 \mathcal{A} \mathcal{B} + 3 \delta_m (\mathcal{A}')^{2} \right)} 
     \Big[ 
    12 \mathcal{A}^{2} \mathcal{A}^{\prime} \left(\mathcal{A} \tfrac{\mathcal{V}^{\prime}}{\mathcal{V}} - 2 \mathcal{A}^{\prime} \right) 
    + 4 \mathcal{A} \mathcal{A}^{\prime} z \left(3 \mathcal{A} \mathcal{A}^{\prime} \tfrac{\mathcal{V}^{\prime}}{\mathcal{V}} + 4 \mathcal{A} \mathcal{B} - 6 \left(\mathcal{A}^{\prime}\right)^{2} \right) 
    \nonumber \\
     & \qquad \qquad \qquad \qquad \qquad \qquad
    - 2 \mathcal{A} z^{2} \left(2 \mathcal{A} \mathcal{A}^{\prime\prime} \mathcal{B} + \mathcal{A} \mathcal{A}^{\prime} \mathcal{B} \tfrac{\mathcal{V}^{\prime}}{\mathcal{V}} - \mathcal{A} \mathcal{A}^{\prime} \mathcal{B}^{\prime} \tfrac{\mathcal{V}^{\prime}}{\mathcal{V}} + 2 \mathcal{A} \mathcal{B}^{2} - 3 \left(\mathcal{A}^{\prime}\right)^{2} \mathcal{B} \right)
    \nonumber \\
     & \qquad \qquad \qquad \qquad \qquad \qquad
    + 6 \delta_{m} \mathcal{A} \left(\mathcal{A}^{\prime}\right)^{2} z \left(4 \mathcal{A}^{\prime} - \mathcal{B} z\right) 
    + 3 \delta_{P} \left(\mathcal{A}^{\prime}\right)^{2} z^{2} \left(\mathcal{A} \mathcal{A}^{\prime} \tfrac{\mathcal{V}^{\prime}}{\mathcal{V}} + 2 \mathcal{A} \mathcal{B} - 2 \left(\mathcal{A}^{\prime}\right)^{2} \right) 
  \Big] \,.
  \label{eq:weff:dynsys}
\end{align}
Here, we have used the cosmological equations to eliminate $H$.
The condition for regular (non-asymptotic) fixed points is given by $\frac{\DD\phi}{\DD N}=z=0$, which means that they must occur at $(\phi_{FP}, 0)$, where $\phi_{FP}$ can be found from the condition~\eqref{eq:general_fixed_point_condition}:
\begin{align}
\label{eq:FP condition:dynsys}
     \tfrac{\mathcal{A}}{\left(2 \mathcal{A} \mathcal{B} + 3 \delta_m (\mathcal{A}')^{2} \right)} \left(\mathcal{A} \tfrac{\mathcal{V}^{\prime}}{\mathcal{V}} - 2 \mathcal{A}^{\prime} \right) \Big|_{FP}  = 0 \,.
\end{align}
We can see that the fixed points coincide in the metric and Palatini cases. From Eq.\ \eqref{eq:weff:dynsys}, we can see that the fixed points are characterized by de Sitter expansion, i.e. $\weff=-1$, in line with the relation between the cosmological evolution of scalar fields and renormalization group equations \cite{Binetruy:2014zya, Pieroni:2015cma}, which also extends to models formulated in Palatini \cite{Karam:2021wzz}.

The stability of the fixed points can be assessed as usual by linearizing the system near the fixed point, which in this case returns
\begin{align}
    \left[ \begin{matrix}  \frac{\DD}{\DD N} \delta \phi\\ \frac{\DD}{\DD N} \delta z\end{matrix} \right] =& 
    \left[ \begin{matrix} 0 & 1 \\ \frac{\partial }{\partial \Phi}\frac{\DD z}{\DD N} & \frac{\partial }{\partial z}\frac{\DD z}{\DD N} \end{matrix} \right]_{FP} \, \,
    \left[ \begin{matrix} \delta \phi \\ \delta z \end{matrix} \right] = \left[\begin{matrix}0 & 1\\- \frac{6 \left(\mathcal{A}^{2} \mathcal{V}^{\prime\prime} - 2 \mathcal{A} \mathcal{A}^{\prime\prime} \mathcal{V} - 2 \left(\mathcal{A}^{\prime}\right)^{2} \mathcal{V}\right)}{\mathcal{V} \left(3 \delta_{m} \left(\mathcal{A}^{\prime}\right)^{2} + 2 \mathcal{A} \mathcal{B}\right)} & -3\end{matrix}\right]_{FP} \left[ \begin{matrix} \delta \phi \\ \delta z \end{matrix} \right],
\end{align}
where $\delta \phi$, $\delta z$ are small perturbations around the fixed point coordinates $(\phi_{FP}, 0)$, the function $\frac{\DD z}{\DD N}$ is the RHS of \eqref{eq:dynsys:z}, and the functions must be evaluated at the fixed point location. A fixed point is stable if the real parts of the eigenvalues of the linearization matrix are negative. The eigenvalues are calculated as
\begin{align}
\label{eq: fixed point eigenvalues}
    \lambda_\pm =& - \frac{3}{2} \pm \frac{3}{2} \sqrt{1 - \frac{8 \left(\mathcal{A}^{2} \mathcal{V}^{\prime\prime} - 2 \mathcal{A} \mathcal{A}^{\prime\prime} \mathcal{V} - 2 \left(\mathcal{A}^{\prime}\right)^{2} \mathcal{V}\right)}{3 \mathcal{V} \left(3 \delta_{m} \left(\mathcal{A}^{\prime}\right)^{2} + 2 \mathcal{A} \mathcal{B}\right)}  } \Bigg|_{FP} \,,
\end{align}
and they are negative when
\begin{align}
\label{eq: fixed point stability condition}
    \frac{\mathcal{A}^{2} \mathcal{V}^{\prime\prime} - 2 \mathcal{A} \mathcal{A}^{\prime\prime} \mathcal{V} - 2 \left(\mathcal{A}^{\prime}\right)^{2} \mathcal{V}}{\mathcal{V} \left(3 \delta_{m} \left(\mathcal{A}^{\prime}\right)^{2} + 2 \mathcal{A} \mathcal{B}\right)} \Big|_{FP} >0 \,,
\end{align}
which matches the condition $\tfrac{1}{\meff}\tfrac{\Veff''}{V} >0$ \cite{Jarv:2021ehj}. Otherwise, when the expression \eqref{eq: fixed point stability condition} is negative, then the fixed point is a saddle with one unstable eigendirection. If the expression \eqref{eq: fixed point stability condition} turns out to be zero, then the eigenvalue $\lambda_+$ is zero as well, and the fixed point is \emph{nonhyperbolic}. To fully analyze a non-hyperbolic point, we would need to go beyond the linear approximation. However, it is often the case that we can read off whether the corresponding eigendirection is stable or not from the phase portrait. In fact, the eigenvalues \eqref{eq: fixed point eigenvalues} can be also computed from the scalar field invariant equation \eqref{Phi_cosmo_equation_in_terms_of_invariants} and written out in terms of the invariants \cite{Jarv:2014hma}. As said before, the fixed points and their stability properties are an invariant feature of scalar-tensor cosmology, irrespective of the chosen frame and parametrization.

The slow roll curve in the phase space is given by  \eqref{eq:slowroll}:
\begin{align}
\label{eq:slowroll:dynsys}
    z = - \frac{2 \mathcal{A} \left(\mathcal{A} \tfrac{\mathcal{V}^{\prime}}{\mathcal{V}} - 2 \mathcal{A}^{\prime} \right)}{\left(2 \mathcal{A} \mathcal{B} + 3 \delta_m \left(\mathcal{A}^{\prime}\right)^{2}\right)} \,.
\end{align}
It satisfies the fixed point condition \eqref{eq:FP condition:dynsys} at $z=0$, and thus we can expect the curve marked by the slow roll condition to start or end at the fixed point. While the fixed point locations \eqref{eq:FP condition:dynsys} and stability properties \eqref{eq: fixed point stability condition} coincide in the metric and Palatini cases due to the no ghost condition \eqref{eq: no ghost condition}, the position of the slow roll approximation curve in the phase space starts to depend on the formalism. The difference will typically increase the further the solutions roll from the de Sitter fixed point, being the greatest at the end of inflationary accelerated expansion. Let us also stress that the slow roll curve \eqref{eq:slowroll:dynsys} itself is not a solution of the full nonlinear system, but just a linear approximation of a trajectory that either leaves the fixed point along an unstable eigendirection or enters along a stable eigendirection.

It is straightforward to rewrite the bound on the physically acceptable values of the variables \eqref{eq:Friedmann_bound} in terms of the dynamical variables as
\begin{align}
\label{eq:Friedmann_bound:dynsys}
    1 + \frac{\mathcal{A}^\prime z}{ \mathcal{A} } - \frac{ \left(2\mathcal{A} \mathcal{B} - 3 \delta_P (\mathcal{A}')^2\right) z^2 }{ 12\mathcal{A}^2 } \geq 0 \,.
\end{align}
We thus see that the physical phase space area is different between the metric formalism and Palatini formalism. For a viable model, the heteroclinic orbit responsible for slow roll inflation must lie in the physical part of the phase space. Thus, even when the values of the scalar field at the fixed points coincide in the metric and Palatini formalisms, we must still ensure that in both cases these points reside within the physical region (explicit examples how this condition can be violated were encountered in Ref.\ \cite{Jarv:2021ehj}).

The behavior of the system at $\phi$ infinity can be revealed with the help of the Poincar\'e compactification
\begin{align}
\label{eq: Poincare variables}
\phi_p = \frac{\phi}{\sqrt{1+\phi^2+z^2}}\,, \qquad
z_p = \frac{z}{\sqrt{1+\phi^2+z^2}} \,.
\end{align}
The inverse relation between the compact variables $\phi_p, z_p$ and the original ones \eqref{eq:dynamical variables} is
\begin{align}
\label{eq: Poincare variables inverse}
\phi=\frac{\phi_p}{\sqrt{1-\phi_p^2-z_p^2}} \,, \qquad z=\frac{z_p}{\sqrt{1-\phi_p^2-z_p^2}} \,.
\end{align}
By construction, the compact variables are bounded, 
$\phi_p^2+z_p^2\leq 1$, and they map an infinite phase plane onto a disc with unit radius. Thus we can rewrite the dynamical system as
\begin{subequations}
\label{eq:dynsys:infinite}
\begin{align}
     \frac{\DD \phi_p}{\DD N} =& \frac{\partial \phi_p}{\partial \phi} \frac{\DD \phi}{\DD N} + \frac{\partial \phi_p}{\partial z} \frac{\DD z}{\DD N} \\
     \frac{\DD z_p}{\DD N} =& \frac{\partial z_p}{\partial \phi} \frac{\DD\phi}{\DD N} + \frac{\partial z_p}{\partial z} \frac{\DD z}{\DD N} 
\end{align}
\end{subequations}
where $\tfrac{\DD \phi}{\DD N}$ and $\tfrac{\DD z}{\DD N}$ are the RHS of \eqref{eq:dynsys} with the variables \eqref{eq: Poincare variables inverse} substituted in. Similarly, it is possible to express the fixed points, slow roll curves, etc.\ in terms these new variables, to get a compact global picture of the full phase space including the asymptotic regions.
In the global picture, most of the solutions originate (or sometimes terminate) at the fixed points which reside where the boundary of the physical phase space area reaches asymptotics. As the physical phase space condition is different in the metric and Palatini formalisms, the state of the universe corresponding to the asymptotic past (or sometimes to the asymptotic future) will be different.
 
\section{Example of Identical actions that yield the same observational parameters in metric and Palatini: Coleman-Weinberg Quartic Potential}
\label{sec:identicalactionsidenticalobs}

In order to get a better feel of how the scalar-tensor models which have proportional invariant potentials and thus predict the same values for the inflationary observables could differ in their overall dynamics, let us first consider a class with identical metric and Palatini actions as an example.

\subsection{Induced gravity Coleman-Weinberg inflation}

As noted in Refs.\ \cite{Racioppi:2017spw,Racioppi:2018zoy} and explained in Ref.\ \cite{Jarv:2020qqm} the scalar field model with quartic potential augmented by Coleman-Weinberg radiative corrections and endowed with a quadratic coupling to gravity,
\begin{align}
\label{eq: CW model}
    \mathcal{A} &= \xi \Phi^2 \,, \qquad
    \mathcal{B} = 1 \,, \qquad
    \mathcal{V} = \frac{\lambda}{4} \left( \ln \left| \frac{\Phi}{\Phi_0} \right| \right)^2 \Phi^4 + \Lambda \,,
\end{align}
yields the same inflationary observables in both metric and Palatini formalisms. Here $\xi$ and $\lambda$ are dimensionless parameters characterizing the model, while $\Phi_0$ can be called the vacuum expectation value of $\Phi$. The cosmological constant $\Lambda$ is a very small quantity compared to the typical scale of inflation, but like in Ref.\ \cite{Jarv:2021qpp} we keep it in the calculations as a regularizing parameter to avoid accidental division by zero in the dynamical system, and apply the limit $\Lambda \to 0$ in the end to present the results. (Without an extra mention, the same procedure applies to all the subsequent examples as well.) 

The potential $\mathcal{V}$ in \eqref{eq: CW model} has stationary points of $\Phi$ at $0$, $\pm \tfrac{\Phi_{0}}{\sqrt{e}}$, and $\pm \Phi_0$. Here the second value corresponds to a local maximum, while the first and last values represent degenerate minima useful in the context of multiple point criticality principle \cite{Racioppi:2021ynx}. However, with the nontrivial coupling function $\mathcal{A}$ the dynamics is generally ruled by the effective potential \eqref{eq: Veff}, which in the limit $\Lambda \to 0$ diverges at $\Phi=0$, has minima at $\Phi=\pm \Phi_0$, and reaches a plateau at infinity. The late time low energy physics takes place around the minimum and we can define the Planck mass as
\begin{align}
\label{eq: CW Planck mass}
    \MP^2 = 8 \pi \xi \Phi_0^2 \,.
\end{align}

\subsection{Inflationary observables}
\label{subsec: CW observables}
 
To get a prediction for the observables, let us quickly go through the algorithm presented in Sec.\ \ref{sec:observables}. The invariant field \eqref{I_Phi} stemming from the model \eqref{eq: CW model} is
\begin{align}
\label{eq: CW IP}
    \IP &= \sqrt{\frac{1 + 6 \xi \delta_m}{\xi}} \, \ln\left|\frac{\Phi}{\Phi_{0}} \right| \,.
 \end{align}   
Note that the original field maps twice onto the invariant field \eqref{eq: CW IP}. The range $\Phi=(0,\Phi_0]$ maps to $\IP=(-\infty,0]$, while $\Phi=[\Phi_0,\infty)$ maps to $\IP=[0,\infty)$. In mirror, also $\Phi=(0,-\Phi_0]$ maps to $\IP=(-\infty,0]$, while $\Phi=[-\Phi_0,-\infty)$ maps to $\IP=[0,\infty)$. At $\Phi_0$ the effective mass \eqref{eq: m_eff} diverges and it is not possible for the solutions to cross over from positive to negative $\Phi$. In terms of the invariant field, this is reflected in the ``distance'' in field space, as the would-be cross-over point is projected away to infinity.

The corresponding invariant potential \eqref{I_V} turns out to be 
\begin{equation}
\label{eq: CW IV}
    \IV = \frac{\lambda}{4\xi(1 + 6 \xi \delta_m)} \IP^2 \,,
\end{equation} 
(cf.\ \cite{Jarv:2016sow} for the metric case). It is the special feature of the model \eqref{eq: CW model} that the invariant potential has the same functional form in the metric and Palatini cases, and only the overall multiplicative constant bears a difference. 
The invariant potential \eqref{eq: CW IV} is analogous to chaotic inflation, and following the recipe of Sec.\ \ref{sec:observables} it is easy to find the slow-roll parameters \eqref{epsilon} and \eqref{eta} to be
$
    \hat{\epsilon} = \hat{\eta} = 2/\mathcal{I}^{2}_{\Phi} \,.
$
In the usual procedure $\hat{\epsilon}=1$ is taken to mark the end of inflation, this leads to 
$
    \IP^\text{end} = \pm \sqrt{2} \,.
$
The number of e-folds \eqref{number.of.efolds} is given by
\begin{equation}
    \widehat{N}{}^{\text{end}}_{*} = \dfrac{1}{4} \left(\mathcal{I}_{\Phi^*}^{2}  - (\mathcal{I}_{\Phi}^{\text{end}})^{2} \right) \,,
\end{equation}
which fixes the inflationary initial value according to the desired number of e-folds as
\begin{align}
\label{eq: CW I_Phi_*}
    \mathcal{I}_\Phi^* &= \pm \left\lbrace \begin{tabular}{ll}
        $14.07$ \,, & \quad $\widehat{N}{}^{\text{end}}_{*} = 50$ \\
        $15.56$ \,, & \quad $\widehat{N}{}^{\text{end}}_{*} = 60$
    \end{tabular} \right. 
\end{align}
From this, we can calculate the expected values of inflationary observables spectral tilt \eqref{ns} and tensor-to-scalar ratio \eqref{r}
\begin{align}
\label{eq: CW n_s}
    n_s =  1-\frac{8}{(\IP^{*})^2} &= \left\lbrace \begin{tabular}{ll}
        0.9596 \,, & \quad  $\widehat{N}{}^{\text{end}}_{*} = 50$ \\
        0.9670 \,, & \quad  $\widehat{N}{}^{\text{end}}_{*} = 60$
    \end{tabular} \right. 
\\
\label{eq: CW r}
    r = \frac{32}{(\IP^{*})^2} &= \left\lbrace \begin{tabular}{ll}
        0.16 \,, & \quad  $\widehat{N}{}^{\text{end}}_{*} = 50$ \\
        0.13 \,, & \quad  $\widehat{N}{}^{\text{end}}_{*} = 60$
    \end{tabular} \right. 
\end{align} 
as well as the scalar amplitude \eqref{scalar.amplitude} 
\begin{equation}
\label{eq: CW A_s }
A_s = \frac{1}{48 \pi^2} \frac{\lambda}{4\xi(1 + 6 \xi \delta_m)} (\IP^{*})^4 \,.
\end{equation}
The latter can be used to estimate the order of magnitude of the parameters
\begin{align}
\label{eq: CW A_s values}
   \frac{\lambda}{4\xi(1 + 6 \xi \delta_m)} = \frac{48 \pi^2 A_s}{\mathcal{I}_{\Phi^*}^{4}} = 
 \begin{cases}
        2.53 \times 10^{-11} \,, & \widehat{N}{}^{\text{end}}_{*} = 50 \\
        1.70 \times 10^{-11} \,, & \widehat{N}{}^{\text{end}}_{*} = 60
    \end{cases} \,.
\end{align}  

The predicted observables $n_s$ and $r$ coincide in the metric and Palatini case, but the scalar amplitude involves a difference. If we take $\lambda \approx 0.129$ as in the quartic (Higgs) potential of the Standard Model, then in the metric case we estimate $\xi_m \sim 10^{4}$ while in the Palatini case $\xi_P \sim 10^{9}$. From the relation \eqref{eq: CW IP} it will then be straightforward to find the field values $\Phi^*$ and $\Phi^{\text{end}}$ corresponding to the start and end of inflation.
Due to the double mapping between the invariant field $\IP$ and the field $\Phi$ in the original parametrization, as explained after Eq.\ \eqref{eq: CW IP}, the inflationary epoch can take place at small field (corresponding to the ``$-$'' sign in $\IP^\text{end} = \pm \sqrt{2}$ and \eqref{eq: CW I_Phi_*}, or large field, corresponding to the ``$+$'' sign respectively, slowly rolling down to $\Phi_0$. 
Since $A_s$ forces the Palatini nonminimal coupling $\xi_P$ to be much bigger than the metric $\xi_m$, the corresponding Palatini field values for $\Phi^*$ and $\Phi^{\text{end}}$ are quite extreme. Because the tensor to scalar ratio \eqref{eq: CW r} is bigger than the current bound \eqref{eq: Planck r} set by Planck, this model is not very realistic in the face of observations. Nevertheless, despite the metric and Palatini models giving the same predictions for $n_s$ and $r$, the values of $\Phi^*$ (in Planck units) at which these predictions are computed can differ from each other by many orders of magnitude.

\subsection{Phase portrait}
\label{subsec: CW phase portrait}

Let us now study this model as a dynamical system in detail, and go through all steps of the method outlined in Sec.\ \ref{sec:dynsys}. 
The calculations for the later examples will proceed in a similar way, but then we can skip some of the technicalities and focus on the key results only. 

In terms of the variables \eqref{eq:dynamical variables} the field equations of the model \eqref{eq: CW model}  written as a dynamical system \eqref{eq:dynsys} read (in the $\Lambda \to 0$ limit) 
\begin{subequations}
\label{eq:dynsys:CW:finite}
\begin{align}
     \phi' =& z \,, \\
     z' =& \frac{\left(\phi + z\right) \left(z^{2} - 6 \phi^{2} \xi - 12 \phi \xi z \right)}{\phi^{2} \left(1+ 6 \delta_{m} \xi\right) \ln{\left(\left|{\frac{\phi}{\phi_{0}}}\right| \right)}} + \frac{z \left(z^{2} - 6 \phi^{2} \xi - 10 \phi \xi z \right)}{2 \phi^{2} \xi \left(1+ 6 \delta_{m} \xi \right)}  + \frac{3 \delta_{m} z \left(z^{2} - 6 \phi^{2} \xi - 10 \phi \xi z \right)}{\phi^{2} \left(1+ 6 \delta_{m} \xi\right)} 
     - \frac{3 \delta_{P} z^{2} \left(2 \phi \xi + 2 \xi z + z \ln{\left(\left|{\frac{\phi}{\phi_{0}}}\right| \right)}\right)}{\phi^{2} \left(1+ 6 \delta_{m} \xi \right) \ln{\left(\left|{\frac{\phi}{\phi_{0}}}\right| \right)}} \,,
\label{eq:dynsys:CW:finite:z}
\end{align}
\end{subequations}
where $\phi_0=\tfrac{\Phi_0}{M}$. Like the original equations, the system contains terms which are common for both metric and Palatini cases, and terms which are different in the two formalisms. Thus very obviously, the dynamics of the two cases is different, as at an arbitrary point in the phase space the flow direction will be different. We can also write out the effective barotropic index \eqref{eq:weff:dynsys}
\begin{align}
\label{eq:weff:CW:finite}
    \weff =& -1 + \frac{2 \left(- 6 \phi^{2} \xi - 12 \phi \xi z + z^{2}\right)}{3 \phi^{2} \left(6 \delta_{m} \xi + 1\right) \ln{\left(\left|{\frac{\phi}{\phi_{0}}}\right| \right)}} + \frac{z \left(- 8 \phi \xi + z\right)}{3 \phi^{2} \xi \left(6 \delta_{m} \xi + 1\right)} 
    + \frac{2 \delta_{m} z \left(- 8 \phi \xi + z\right)}{\phi^{2} \left(6 \delta_{m} \xi + 1\right)} - \frac{2 \delta_{P} z^{2} \left(2 \xi + \ln{\left(\left|{\frac{\phi}{\phi_{0}}}\right| \right)}\right)}{\phi^{2} \left(6 \delta_{m} \xi + 1\right) \ln{\left(\left|{\frac{\phi}{\phi_{0}}}\right| \right)}} \,,
\end{align}
the bound on the physical phase space \eqref{eq:Friedmann_bound:dynsys},
\begin{align}
\label{eq:Friedmann_bound:CW}
    \frac{6 \phi^{2} \xi + 12 \phi \xi z + 6 \delta_{P} \xi z^{2} - z^{2}}{6 \phi^{2} \xi} \geq 0 \,,
\end{align}
and the slow roll curve \eqref{eq:slowroll:dynsys},
\begin{align}
\label{eq:slowroll:CW:finite}
    z =& - \frac{2 \phi \xi }{ \left(6 \delta_{m} \xi + 1\right) \ln{\left(\left|{\frac{\phi}{\phi_{0}}}\right| \right)}}
\end{align}
(all in the $\Lambda \to 0$ limit). All these expressions are different in the metric vs.\ Palatini case. For instance the Friedmann constraint \eqref{eq:Friedmann_bound:CW} stipulates the allowed values of $z$ to reside in a single band for all $\xi$ in the metric formalism, while in the Palatini formalism a single band is allowed for small $\xi$ but two bands separated by a restricted zone appear for larger $\xi$:
\begin{subequations}
\label{eq: CW bound on phi}
\begin{align}
    \textrm{metric } 0 < \xi : & \qquad - \sqrt{6\xi(1+6\xi)} \phi \leq z \leq \sqrt{6\xi(1+6\xi)} \phi \,, \\
    \textrm{Palatini } 0 < \xi \leq \tfrac{1}{6} : & \qquad -\frac{\left(\sqrt{6 \xi}-6\xi \mathrm{sign}(\phi) \right)\phi \, \mathrm{sign}(\phi)}{1-6\xi} \leq z \leq \frac{\left(\sqrt{6 \xi}+6\xi \mathrm{sign}(\phi) \right)\phi \, \mathrm{sign}(\phi)}{1-6\xi} \,, \\
    \textrm{Palatini } \tfrac{1}{6} < \xi : & \qquad  z \leq -\frac{\left(\sqrt{6 \xi}+6\xi \mathrm{sign}(\phi) \right)\phi \, \mathrm{sign}(\phi)}{6\xi-1} \,, \quad \frac{\left(\sqrt{6 \xi}-6\xi \mathrm{sign}(\phi) \right)\phi \, \mathrm{sign}(\phi)}{6\xi-1} \leq z \,,
\end{align}
\end{subequations}
Hence the layout of the phase space is qualitatively different for the metric and Palatini versions. It is easy to check that the effective barotropic index \eqref{eq:weff:CW:finite} varies along the boundary of the physical phase space taking different values for the metric and Palatini cases. In particular, on the boundaries \eqref{eq:Friedmann_bound:CW} in the asymptotic limit $|\phi| \to \infty$ the effective barotropic index is
\begin{align}
\label{eq:weff b CW}
    \weff{}_b^\pm =& \pm \frac{4 \sqrt{6 \xi} \sqrt{1+6\delta_m\xi}}{3(1-6\delta_P\xi )} + \frac{1+2\xi(1+3\delta_m)}{1-6\delta_P\xi} \,.
\end{align}
These limiting values are important in describing states of the asymptotic fixed points.

Despite these differences the fixed point condition~\eqref{eq:FP condition:dynsys} is satisfied (by applying the limit $\Lambda \to 0$ in the last step) at $(\phi,z)$
\begin{align}
    A_\pm &: \left(\pm \phi_0, 0 \right) \,, \\
    B &: (0,0) \,. 
\end{align}
Thus the fixed points coincide in the metric and Palatini case. Comparison with the expression \eqref{eq: fixed point stability condition} reveals that in both the metric and Palatini version the points at $\phi=\pm \phi_0$ are stable, while the point at $\phi=0$ is nonhyperbolic. Hence the former points would correspond to a late universe where the solution trajectories end up, while the latter point can correspond to an early stage of the universe, from where the solutions depart. The coordinates of the second point satisfy the slow roll condition \eqref{eq:slowroll:CW:finite}, therefore it can serve as a launching pad for slow roll motion.

To see the full global picture of the system, we need to transform into the compact Poincar\'e variables \eqref{eq: Poincare variables}. In terms of these the system \eqref{eq:dynsys:CW:finite} can be rewritten like \eqref{eq:dynsys:infinite} as
\begin{subequations}
\label{eq:dynsys:CW:infinite}
\begin{align}
     \phi_p' =& \frac{z_{p} \left(6 \phi_{p}^{2} \xi z_{p} + 2 \phi_{p} \xi q^{2} + 12 \phi_{p} \xi z_{p}^{2} - z_{p}^{3}\right)}{2 \phi_{p} \xi \left(6 \delta_{m} \xi + 1\right)} 
     \frac{z_{p} \left(\phi_{p} + z_{p}\right) \left(6 \phi_{p}^{2} \xi + 12 \phi_{p} \xi z_{p} - z_{p}^{2}\right)}{\phi_{p} \left(6 \delta_{m} \xi + 1\right) \ln{\left(\frac{\left|{\frac{\phi_{p}}{\phi_{0}}}\right|}{q} \right)}} \nonumber \\
     & + \frac{3 \delta_{m} z_{p} \left(6 \phi_{p}^{2} \xi z_{p} + 2 \phi_{p} \xi q^{2} + 12 \phi_{p} \xi z_{p}^{2} - z_{p}^{3}\right)}{\phi_{p} \left(6 \delta_{m} \xi + 1\right)} 
     +\frac{3 \delta_{P} z_{p}^{3} \left(2 \phi_{p} \xi + 2 \xi z_{p} + z_{p} \ln{\left(\frac{\left|{\frac{\phi_{p}}{\phi_{0}}}\right|}{q} \right)}\right)}{\phi_{p} \left(6 \delta_{m} \xi + 1\right) \ln{\left(\frac{\left|{\frac{\phi_{p}}{\phi_{0}}}\right|}{q} \right)}}\\
     z_p' =& - \frac{z_{p} \left(6 \phi_{p}^{4} \xi + 12 \phi_{p}^{3} \xi z_{p} + 6 \phi_{p}^{2} \xi q^{2} - \phi_{p}^{2} z_{p}^{2} + 10 \phi_{p} \xi q^{2} z_{p} - q^{2} z_{p}^{2}\right)}{2 \phi_{p}^{2} \xi \left(6 \delta_{m} \xi + 1\right)} 
     - \frac{\left(\phi_{p} + z_{p}\right) \left(\phi_{p}^{2} + q^{2}\right) \left(6 \phi_{p}^{2} \xi + 12 \phi_{p} \xi z_{p} - z_{p}^{2}\right)}{\phi_{p}^{2} \left(6 \delta_{m} \xi + 1\right) \ln{\left(\frac{\left|{\frac{\phi_{p}}{\phi_{0}}}\right|}{q} \right)}} \nonumber \\ 
     & - \frac{3 \delta_{m} z_{p} \left(6 \phi_{p}^{4} \xi + 12 \phi_{p}^{3} \xi z_{p} + 6 \phi_{p}^{2} \xi q^{2} - \phi_{p}^{2} z_{p}^{2} + 10 \phi_{p} \xi q^{2} z_{p} - q^{2} z_{p}^{2}\right)}{\phi_{p}^{2} \left(6 \delta_{m} \xi + 1\right)} 
     - \frac{3 \delta_{P} z_{p}^{2} \left(\phi_{p}^{2} + q^{2}\right) \left(2 \phi_{p} \xi + 2 \xi z_{p} + z_{p} \ln{\left(\frac{\left|{\frac{\phi_{p}}{\phi_{0}}}\right|}{q} \right)}\right)}{\phi_{p}^{2} \left(6 \delta_{m} \xi + 1\right) \ln{\left(\frac{\left|{\frac{\phi_{p}}{\phi_{0}}}\right|}{q} \right)}} 
\end{align}
\end{subequations}
where $q^2=1-\phi_p^2-z_p^2$ is a shorthand notation. In these variables the effective barotropic index is
\begin{align}
\label{eq:weff:CW:infinite}
    \weff =& - 1 - \frac{z_{p} \left(8 \phi_{p} \xi - z_{p}\right)}{3 \phi_{p}^{2} \xi \left(6 \delta_{m} \xi + 1\right)}
    - \frac{2 \left(6 \phi_{p}^{2} \xi + 12 \phi_{p} \xi z_{p} - z_{p}^{2}\right)}{3 \phi_{p}^{2} \left(6 \delta_{m} \xi + 1\right) \ln{\left(\frac{\left|{\frac{\phi_{p}}{\phi_{0}}}\right|}{q} \right)}} 
     - \frac{2 \delta_{m} z_{p} \left(8 \phi_{p} \xi - z_{p}\right)}{\phi_{p}^{2} \left(6 \delta_{m} \xi + 1\right)} 
     - \frac{2 \delta_{P} z_{p}^{2} \left(2 \xi + \ln{\left(\frac{\left|{\frac{\phi_{p}}{\phi_{0}}}\right|}{q} \right)}\right)}{\phi_{p}^{2} \left(6 \delta_{m} \xi + 1\right) \ln{\left(\frac{\left|{\frac{\phi_{p}}{\phi_{0}}}\right|}{q} \right)}} \,,
\end{align}
the bound determining the physical phase space is given by
\begin{align}
\label{eq:Friedmann_bound:CW:infinite}
    \frac{6 \delta_{P} \xi z_{p}^{2} + 6 \phi_{p}^{2} \xi + 12 \phi_{p} \xi z_{p} - z_{p}^{2}}{6 \phi_{p}^{2} \xi} \geq 0 \,,
\end{align}
and the slow roll approximation reads as
\begin{align}
\label{eq:slowrollCW:infinite}
    z_p =& - \frac{2 \phi_{p} \xi }{\left(6 \delta_{m} \xi + 1\right) \ln{\left(\frac{\left|{\frac{\phi_{p}}{\phi_{0}}}\right|}{q} \right)}} \,.
\end{align}
Similarly, in the coordinates $(\phi_p, z_p)$ the fixed points map to
\begin{align}
    A_\pm &: \left(\pm \frac{\phi_0}{\sqrt{1+\phi_0^2}}, 0 \right) \,, \\
    B &: (0,0) \,.
\end{align}
and retain their properties.

There is an exact correspondence between all the points of the finite phase space and the compactified global phase space. By the compactification, the main new information we get is about the behavior of the system in the asymptotics. For example, we can go to the limit $\phi_p \to \pm\sqrt{1-z_p^2}$ whereby the system \eqref{eq:dynsys:CW:infinite} on the asymptotic 1-dimensional circumference of the unit disc reduces to
\begin{subequations}
\label{eq:dynsys:CW:infinite:asymptotic}
\begin{align}
     z_p' =& \mp 6 z_{p}^{2} \sqrt{1 - z_{p}^{2}} + \frac{z_{p} \left(6 \xi z_{p}^{2} - 6 \xi + z_{p}^{2}\right)}{2 \xi \left(6 \delta_{m} \xi + 1\right)} + \frac{3 \delta_{m} z_{p} \left(6 \xi z_{p}^{2} - 6 \xi + z_{p}^{2}\right)}{6 \delta_{m} \xi + 1} - \frac{3 \delta_{P} z_{p}^{3}}{6 \delta_{m} \xi + 1} \,.
\end{align}
\end{subequations}
Here the upper (lower) sign corresponds to the ``right'' (``left'') hemisphere $\phi_p>0$ ($\phi_p<0$). This equation has six solutions corresponding to asymptotic fixed points. The first two points are the same for metric and Palatini cases:
\begin{align}
    C_\pm : & (\pm 1, 0 ) \,.
\end{align}
For the 1-dimensional asymptotic flow on the unit disc circumference these points are attractors. However, for the full 2-dimensional flow where the disc interior is included, the other eigenvalue corresponding to an eigendirection towards the interior is zero, and the points $C_\pm$ are nonhyperbolic. Substituting the fixed point coordinates into the expression of the barotropic index \eqref{eq:weff:CW:infinite} shows that the expansion is like de Sitter, since $\weff=-1$. However, one should not jump directly to the conclusion that the point is the usual de Sitter where the scalar field has stopped and the Hubble function $H$ is constant. Rather a more closer analysis reveals that in this asymptotic state the scalar field evolves along with the changing $H$, hence calling it `asymptotic de Sitter' is more appropriate \cite{Skugoreva:2014gka,Jarv:2021qpp}.

The other asymptotic fixed points are
\begin{align}
    D_\pm : (\pm \sqrt{1-Z_-^2}, \mp Z_-) \,, \\
    E_\pm : (\pm \sqrt{1-Z_+^2}, \pm Z_+) \,, 
\end{align}
where
\begin{align}
    Z_\pm =& \frac{\delta_{m} \sqrt{\pm 12 \sqrt{6} \xi^{\frac{3}{2}} \sqrt{6 \xi + 1} + 108 \xi^{2} + 6 \xi}}{\sqrt{180 \xi^{2} + 12 \xi + 1}}
    \mp \frac{\delta_{P} \sqrt{\pm 12 \sqrt{6} \xi^{\frac{3}{2}} + 72 \xi^{2} + 6 \xi}}{\sqrt{144 \xi^{2} + 1}} 
\end{align}
These reside precisely in the limits where the bound on the physical phase space \eqref{eq:Friedmann_bound:CW:infinite} reaches the asymptotics. The points $D_\pm$ are unstable nodes, meaning a host of trajectories start from them. On the other hand, the points $E_\pm$ are saddles with a repulsive eigendirection along the perimeter of the infinity disc, and an attractive eigendirection along the physical phase space boundary. Although the points $D_\pm$ and $E_\pm$ have the same stability features in the metric and Palatini cases, they correspond to a different dynamical regime as far as the spacetime expansion and scalar field evolution is concerned. Namely, the effective barotropic index at these points is given by Eq.\ \eqref{eq:weff b CW} as follows:
\begin{align}
    \weff(D_{\pm}) &= \weff{}_{b}^- \,, \qquad \weff(E_{\pm}) = \weff{}_{b}^+ \,.
\end{align}
From Eq.\ \eqref{eq:weff_to_a(t)} we can find the respective Hubble functions $H(t)$ and the scale factors $a(t)$, and then integrating the Friedmann equation \eqref{mpf:first:friedmann:1}  also the state of the scalar field $\Phi(t)$. In contrast to the points $E_\pm$ which correspond to very particular solutions, the points $D_\pm$ correspond to the generic initial state of large field evolution, and are worth a closer look. In the metric case this state is given by (compare with Refs.\ \cite{Carloni:2007eu,Skugoreva:2014gka,Jarv:2021qpp})
\begin{align}
\label{eq: CW asymptotic metric}
    a & \sim (t-t_0)^{\frac{1}{3+12\xi-2\sqrt{6\xi(1+6\xi)}}} \,, \qquad |\Phi| \sim (t-t_0)^{-\frac{-6\xi+\sqrt{6\xi(1+6\xi)}}{3+12\xi-2\sqrt{6\xi(1+6\xi)}}} \,.
\end{align}
In the Palatini case the respective state is
\begin{align}
\label{eq: CW asymptotic Palatini}
    a & \sim (t-t_0)^{\frac{1-6\xi}{3-2\sqrt{6\xi}}} \,, \qquad |\Phi| \sim (t-t_0)^{-\frac{6\xi-\sqrt{6\xi}}{6\xi+2\sqrt{6\xi}-3}} \,.
\end{align}
In both cases there is an initial singularity at $t_0$ where the scale factor starts growing from zero and the scalar field begins at infinite value, rolling quickly down from the infinite potential. This state corresponds to a maximally kinetic regime in the large field limit. In the metric case it is has been found to occur for both asymptotically quadratic and quartic potentials in models characterized by $\mathcal{A}\sim \xi\Phi^2$ for large $\Phi$ (cf.\ Refs.\ \cite{Carloni:2007eu,Jarv:2021qpp}). In the Palatini case, the behavior is qualitatively the same but with a different power dependence on the parameter $\xi$.

\begin{figure*}
	\centering
	\subfigure[]{
		\includegraphics[width=0.45\textwidth]{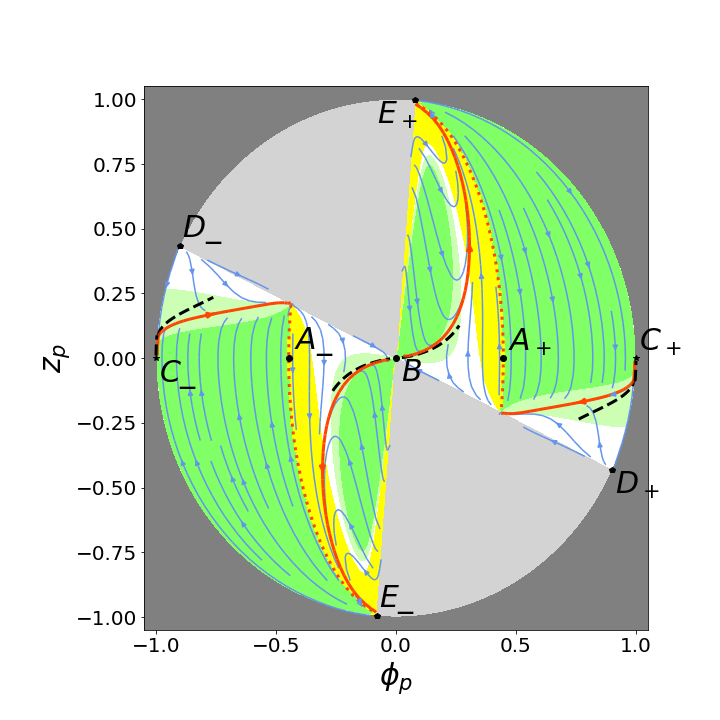} \label{fig:metricCWinfinite}}
	\subfigure[]{
		\includegraphics[width=0.45\textwidth]{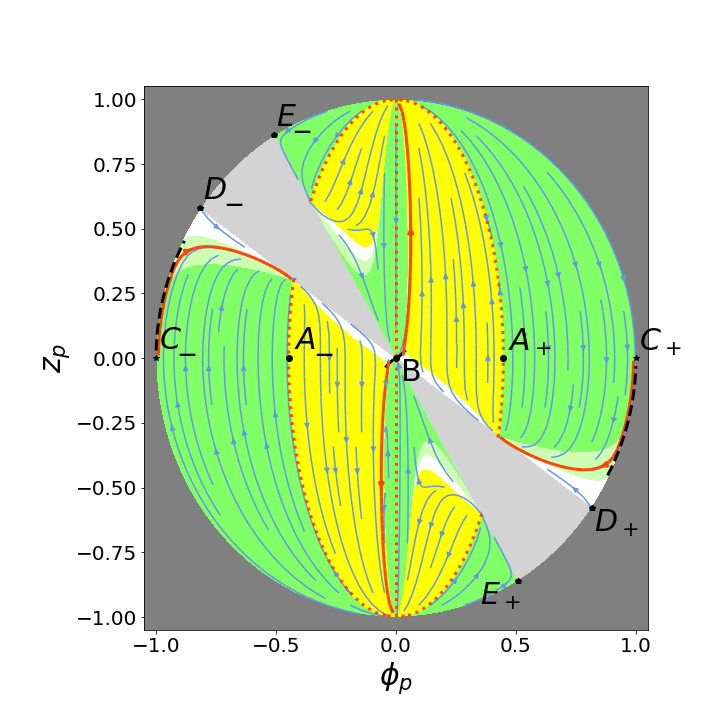} \label{fig:PalatiniCWinfinite}}
  \caption{Cosmological phase portraits of the Coleman-Weinberg model \eqref{eq: CW model} with $\xi=1$, $\lambda=0.129$, $\phi_0=0.5$ ($\Phi_0 \approx 0.199 \, M_{\mathrm{Pl}}$), $\Lambda=0$ in the metric formalism (panel a) and Palatini formalism (panel b). Green background stands for superaccelerated, light green accelerated, white decelerated, and yellow superstiff expansion, while grey covers the unphysical region. Orange trajectories are heteroclinic orbits between the fixed points, and the dashed curve marks the path of slow roll approximation.}
\label{fig:CWmetricandPalatini}
\end{figure*}

With this information, we can assemble a global picture of the dynamics in the models \eqref{eq: CW model}, depicted on Fig.\ \ref{fig:metricCWinfinite} in the metric and on Fig.\ \ref{fig:PalatiniCWinfinite} in the Palatini case. For the plots we have chosen suitable values of the model parameters $\xi$, $\lambda$, and $\Phi_0$ to make the key features of the phase portrait discernible. For other values of the parameters the pictures are qualitatively similar, but some details like the deceleration zone become rather narrow and hard to discern. Here the grey area is ruled out by the Friedmann constraint \eqref{eq:Friedmann_bound:CW:infinite}, while the background coloring of the physically available phase space follows the effective barotropic index \eqref{eq:weff:CW:infinite}: dark green for superaccelerated ($\weff<-1$), light green for accelerated ($-1 \leq \weff < -\tfrac{1}{3}$), white for decelerated ($-\tfrac{1}{3} \leq \weff < 1$), and yellow for ``superstiff'' ($\weff \geq 1$) expansion. The trajectories in blue are some representative solutions. We see that generic trajectories are attracted to the ``master'' solutions (drawn in orange color) which are the heteroclinic orbits flowing from the points $C_\pm$  (which acts as an origin of ``large field'' inflation'')  or $B$ (which acts as an origin of ``small field'' inflation) to the late time attractors $A_\pm$. The slow roll approximations \eqref{eq:slowrollCW:infinite} are drawn as dashed black curves. They start from the fixed points $C_\pm$ or $B$ like the true ``master'' solutions but then gradually diverge.

In the vanishing cosmological constant limit the Eq.\ \eqref{eq:dynsys:CW:finite:z} harbors a singularity at $\phi=\phi_0$. This has the effect of deforming the phase space by ``stretching'' the points $A_\pm$ along the $z$ direction and forcing the nearby trajectories to flow tangent to it, until they reach the point where the singular line touches the physical boundary of the phase space. This is an effect that arises due to the specific choice of the dynamical variables \eqref{eq:dynamical variables} where the dynamical system has terms divided by the potential $\mathcal{V}$, and the zeroes of the potential make that trouble. However, this artifact does not concern our investigation for two reasons. First, as mentioned above, a realistic universe has a small nonvanishing cosmological constant, the potential does not have a true zero, and singularity is not really there. (Sample phase portraits with nonzero cosmological constant can be seen in the appendix of Ref.\ \cite{Jarv:2021qpp}, where the oscillating behavior of the scalar field is clearly visible.) Second, this value pertains to the late universe which is attained only after the process of inflation. The inflationary accelerated expansion occurs on the (orange) ``master'' trajectory until it reaches the deceleration (white) zone. After that point other processes like reheating must be taken into account and the usefulness of the model is exhausted anyway. 

In summary, the induced gravity Coleman-Weinberg model \eqref{eq: CW model} is a specific instance of a theory where the inflationary observables \eqref{eq: CW n_s} and \eqref{eq: CW r} coincide in the metric and Palatini versions. This, however, does not mean that all of the respective dynamics is identical, as the dynamical systems \eqref{eq:dynsys:CW:infinite} still differ from each other in some of the terms. It is a generic property of scalar-tensor gravity that the fixed points and their stability properties are the same in metric and Palatini. Inflation is ruled by a particular ``master'' trajectory that flows out of a saddle-like fixed point to an attractor fixed point, hence if inflation is possible, then both in metric and Palatini cases, and the present Coleman-Weinberg example illustrates that. The other features of the dynamics involve differences, including the asymptotic states \eqref{eq: CW asymptotic metric}, \eqref{eq: CW asymptotic Palatini}, and the boundaries of the physical phase space \eqref{eq: CW bound on phi}. The latter means that while in the metric and $\xi<1/6$ Palatini cases for all initial conditions the dynamics will approach the inflationary ``master'' trajectory, in the $\xi>1/6$ Palatini case there is an extra sector of physically allowed initial conditions where the approach to the ``master'' solution is impossible. This feature illustrates how the global phase portraits provide a useful information. A further difference appears in the ``end'' and ``start'' points of inflation, which we elaborate on in the next subsection.

\subsection{The inflationary trajectory}
\label{subsec: CW inflationary trajectory}

In Sec.\ \ref{subsec: CW observables}, the observables were determined by relying on the invariant (essentially Einstein frame) slow roll approximation. The invariant field value $\IP^{\text{end}}$ at the end inflation was defined by the condition that on the slow roll curve $\hat \epsilon = 1$ or equivalently $\hat w_{\mathrm{eff}}= -1/3$, whereby the expansion in terms of the invariant metric \eqref{FLRW:hat} switches from acceleration to deceleration. The observables \eqref{eq: CW n_s}, \eqref{eq: CW r}, \eqref{eq: CW A_s } were then computed at $\IP^*$, which was obtained by tracing a certain number of invariant e-folds $\widehat{N}$ backwards in time before the end of inflation assuming the slow roll conditions hold. By the mapping \eqref{eq: CW IP} we could find the respective Jordan frame field values $\Phi^{\text{end}}$ and $\Phi^*$.

Our description of the inflationary dynamics directly in the Jordan frame offers an alternative way to ascertain $\Phi^{\text{end}}$ and $\Phi^*$.
On a phase portrait, we can pretty well trace the ``master'' inflationary trajectory (fat orange curve) as well as its slow roll approximation (dashed black line). Since the slow roll curve is given by an analytical expression \eqref{eq:slowroll:CW:finite}, the field value marking the end of inflation can be found by substituting the slow roll condition \eqref{eq:slowroll:CW:finite} into the effective barotropic index \eqref{eq:weff:CW:finite} and setting the end to be the exit from accelerated expansion in the Jordan frame, i.e.\ by solving
\begin{align}
    \weff(\phi^{\text{end}}_{\text{slow roll}}) &= -1 + \frac{4 \xi}{3 \left(6 \xi \delta_m + 1\right) \ln\left(\left|\frac{\phi}{\phi_0}\right|\right)} + \frac{4 \xi \left(6 (1+\delta_m) \xi + 1\right)}{3 \left(6 \xi \delta_m + 1\right)^{2} \ln\left(\left|\frac{\phi}{\phi_0}\right|\right)^{2}} + \frac{8 \xi^{2}(1-6 \xi \delta_P)}{3 \left(6 \xi \delta_m + 1\right)^{3} \ln\left(\left|\frac{\phi}{\phi_0}\right|\right)^{3}} = -\frac{1}{3} \,
\end{align}
for $\phi$. The starting value $\phi^{*}$ can then be determined from the Jordan frame e-folds \eqref{eq: N integral}, where the integral from $\phi^{*}$ to $\phi^{\text{end}}$ runs over the slow roll curve \eqref{eq:slowroll:CW:finite} and gives the required number of e-folds,
\begin{align}
    N{}^{\text{end}}_{*} = \left[-\frac{(1+\delta_m \xi) \ln\left(\left|\frac{\phi}{\phi_0}\right|\right)^2 }{4 \xi} \right]^{\phi^{\text{end}}}_{\phi^{*}} \,.
\end{align}
Finally, conversion from the dimensionless variables \eqref{eq:dynamical variables} to the units in terms of the Planck mass \eqref{eq: CW Planck mass} can be simply carried out by noticing  that
\begin{align}
    \frac{\Phi}{\Phi_0} &= \frac{\phi}{\phi_0} \,,
\end{align}
whereby the arbitrary mass quantity $M$ in \eqref{eq:dynamical variables} cancels out. 

In the metric formalism, the Jordan frame slow roll condition \eqref{eq:slowroll} has been shown to coincide with the Einstein frame (i.e.\ invariant) slow roll condition translated into the Jordan frame \cite{Akin:2020mcr,Jarv:2021qpp,Karciauskas:2022jzd}. In the Palatini case, an extra check is required, but even if given the same path as slow roll, the invariant and Jordan definitions of the end of inflation still correspond to a slightly different point on that path as a moment in the evolution of the universe and hence a slightly different field value $\Phi^{\text{end}}$. As a second step, counting back 50 or 60 e-folds of accelerated expansion on that path can be done either in the invariant (Einstein) or Jordan frame units, which are again slightly different, cf.\ \eqref{eq: N in Jordan and Einstein}, giving slightly different starting values of $\Phi^*$. Since the fixed point structure and their de Sitter nature is a frame invariant feature in scalar-tensor cosmologies, but slow roll is a small deviation from that behavior (see the discussions around Eqs.\ \eqref{eq:slowroll} and \eqref{eq:slowroll:dynsys}), the invariant and Jordan results agree approximately, but not precisely. 

Furthermore, a closely related issue is the goodness of the slow roll approximation in relation to the true inflationary trajectory. As can be observed on Fig.\ \ref{fig:CWmetricandPalatini}, the slow roll curve \eqref{eq:slowroll:CW:finite} (black dashed) and the exact nonlinear ``master'' solution (orange) start from the same fixed points, but begin to deviate from each other as the inflationary behavior nears an end. Since all other trajectories approach very close to the ``master'' solution, it is not hard to numerically follow that trajectory, and ascertain the actual Jordan frame values $\Phi^{\text{end}}$ and $\Phi^{*}$ on the true ``master'' path by catching the moment the trajectory exits the acceleration zone and counting the required number of e-folds backwards in time. That would be yet another possibility to determine the endpoint of inflation \textbf{$\Phi^{\text{end}}$} and the input value $\Phi^{*}$ for the predictions on the observable perturbative spectrum. It is noteworthy, that in both metric and Palatini cases the slow roll approximation predicts the end of inflation at much higher field values than experienced by the true nonlinear solution.

As observations are getting more precise, an obvious question is that of the impact of the difference between alternative definitions of $\Phi^*$ on the actual predictions. This question is closely related to the definition of e-folds: whether they are defined with respect to the expansion of the Universe in the Jordan or Einstein frame is equivalent to choosing a different frame to define $\Phi^*$. Similarly, the end of inflation value $\Phi^{\text{end}}$ has relevance for the preheating and reheating phases in the aftermath of inflation, and different definitions may impact the results of modelling particle generation. The effect of selecting a frame in which $\Phi^*$ is defined is the source of the apparent ``frame dependence'' of the observables  \cite{Postma:2014vaa,Karam:2017zno}: if it is transformed alongside all other quantities, observables are invariant between frames. However, selecting which frame it is going to be originally defined is a physically meaningful choice, especially when differing formalisms of gravity are involved, and one whose investigation we will leave for further work. Here only note in the context of metric vs.\ Palatini models, that both cases allow different definitions as briefly outlined above. Our unifying approach helps to address these issues in both formalisms simultaneously.

\section{Different  actions that yield the same observational parameters in metric and Palatini}
\label{sec:differentactionsidenticalobs}

In this section, we will consider various actions that, to first order at least, return the same observable parameters.  We will begin with Starobinsky inflation, which is supported very well observationally, and further describe other models that reduce to similar representations. We will skip the presentation of the calculations that lead to the global phase portraits, as the method was already illustrated in significant detail in the previous section.

\subsection{Starobinsky inflation in metric formalism}
\label{subsec: metric Starobinsky}

The Starobinsky model \cite{Starobinsky:1980te} as originally proposed is closer to trace-anomaly driven inflation driven by an effective action, but now more commonly refers to its classical analog expressed as an $f(R)$ theory \cite{Vilenkin:1985md}, given by
\begin{align}
\label{eq: Starobinsky lagrangian}
    \mathcal{L}_{f(R)} &= f(R) = M^2 R + \beta R^2,
\end{align}
where the parameter $\beta$ is dimensionless, and $M$ is once again an arbitrary mass parameter, selected later such that the theory reduces to Einstein gravity at the low curvature limit. 
Through a Legendre transform, it can be written in the scalar-tensor form. This is done by introducing a scalar field $M\Phi= M\frac{\DD  f(R)}{\DD R}=M^2+2\beta R$, whereby the dynamically equivalent Lagrangian is expressed in terms of a non-minimally coupled scalar-tensor theory:
\begin{align}
    \mathcal{L}_{\Phi} &= M \Phi R - \frac{M^2}{4 \beta } \left( \Phi - M \right)^2 \,.
\end{align}
We can immediately read off the model functions
\begin{align}
\label{eq: Starobinsky metric model}
    \mathcal{A} &= M \Phi \,, \qquad
    \mathcal{B} = 0 \,, \qquad
    \mathcal{V} = \frac{M^2}{8 \beta } \left( \Phi - M \right)^2 \,.
\end{align}
Here the obvious assumption is that $\Phi$ is positive, otherwise we entertain antigravity. To see the predicted observables we can introduce the invariant field \eqref{I_Phi}:
\begin{align}
 \IP &= \int \, \DD \Phi \sqrt{\frac{3}{2} \frac{M^2}{\Phi^2 M^2}} 
= \sqrt{\frac{3}{2}} \, \ln \left(\frac{\Phi}{\Phi_{0}} \right) \,.   
\end{align}
This definition necessarily includes an arbitrary mass scale (from the integration constant $\Phi_0$) that we cannot set to zero, and thus we set it to the only mass scale present in the theory, i.e. $\Phi_{0} = M$, giving finally
\begin{align}
\label{eq: I_Phi Starobinsky}
    \IP &= \sqrt{\frac{3}{2}} \ln \left(\frac{\Phi}{M} \right).
\end{align}
This can be inverted, and substituted into \eqref{I_V} to get the invariant potential
\begin{align}
\label{eq: I_V Starobinsky}
    \mathcal{I}_{\mathrm{V}} &= \frac{1}{8 \beta} \left(1 - e^{-\sqrt{\tfrac{2}{3}} \mathcal{I}_\Phi} \right)^2.
\end{align}
Positive invariant field \eqref{eq: I_Phi Starobinsky} represents a large original field, $M<\Phi$, whereby the invariant potential \eqref{eq: I_V Starobinsky} flattens to a plateau asymptotically. On the other hand, the negative invariant field \eqref{eq: I_Phi Starobinsky} represents a small original field, $0<\Phi<M$, while the invariant potential \eqref{eq: I_V Starobinsky} grows exponentially as $\IP$ decreases towards more negative values (and $\Phi$ approaches zero). We can only expect inflation to occur for flat invariant potential, hence here at large positive $\IP$, where the field would roll really slowly. In terms of the original field that region is represented by flat effective potential $\Veff$ \eqref{eq: Veff} although the effective mass $\meff$ \eqref{eq: m_eff} is asymptotically zero. The value $\IP=0$  gives the minimum of the invariant potential, which is the destiny of late time evolution. We can check that the corresponding value $\Phi=M$ is an attractive fixed point \eqref{eq:general_fixed_point_condition} of the original field. In general, the model has two parameters, $\beta$ and $M$, only the first of which is dimensionless. We can fix $M = \MP$ such that the theory reduces to general relativity at the low curvature limit of the Starobinsky lagrangian \eqref{eq: Starobinsky lagrangian}. This is also consistent with the late universe where the scalar field stabilizes at an attractive fixed point and $\Acal \to \MP^2$ in \eqref{eq: Starobinsky metric model}. Thus we are left with one free parameter, $\beta$, which is to be fixed through the amplitude $A_s$.

With the invariant formalism, we can find the slow roll parameters \eqref{epsilon}, \eqref{eta} as
\begin{align}
    \hat{\varepsilon} = \frac{4 \, e^{-2 \sqrt{\frac{2}{3}} \IP}}{3 \left(1 - e^{-\sqrt\frac{2}{3} \IP} \right)^2},
\qquad
    \hat{\eta} =  \frac{4 \, e^{-2 \sqrt{\frac{2}{3}} \IP} \left(2  - e^{\sqrt{\frac{2}{3}} \IP} \right)}{3 \left(1 - e^{-\sqrt\frac{2}{3} \IP} \right)^2}.
\end{align}
Calculating the end of inflation at $\hat{\epsilon} =1$ and expressing the number of e-folds in terms of the field through \eqref{number.of.efolds} and then inverting, we can express the value of the invariant field in terms of $N$, which allows us to express
the observables \eqref{ns}, \eqref{r} at the horizon exit (for $\widehat N = 60$) as follows:
\begin{align}
\label{eq: n_s Starobinsky}
    n_{s} &\approx 1 - 6 \hat{\epsilon} + 2 \hat{\eta}  = 0.9656 \,, \\
\label{eq: r Starobinsky}
    r &\approx 16 \hat{\epsilon}  = 0.0034 \,.
\end{align}
This result is in agreement with approximated results (for $\widehat{N}=60$)
\begin{align}
    n_{s} \approx 1 - \frac{2}{\widehat N} = 0.9666 \,, 
\qquad
    r \approx \frac{12}{\widehat N^2}  = 0.0033 \,.
\end{align}
It is further possible to determine the value of $\beta$ by matching to the amplitude of the spectrum \eqref{scalar.amplitude}. Using ${\cal I}_{\rm V}/{\hat \epsilon} = (2.1 \pm 0.059) \times 10^{-7}$ \cite{Planck:2018vyg} (expressed in units of $M=\MP$ which are the natural choice of units for this theory), and evaluating at 60 e-folds returns
\begin{align}
\beta = 1.15^{+0.033}_{-0.032} \times 10^9.
\end{align}
These values of the observables calculated via the invariant formalism agree with previous studies of the model \cite{Mavromatos:2020kzj,DiValentino:2016nni}. It is not ruled out by the constraints from the late universe \cite{Gomes:2023xzk}.

Due to the nature of the phase space variables, neither the dynamical system \eqref{eq:dynsys} nor any of the phase space quantities like $\weff(\phi,z)$ depend on the overall factor $\beta$ of the potential. We will skip the presentation of all the calculational details which follow the same routine as in the previous example Sec.\ \ref{subsec: CW phase portrait}--\ref{subsec: CW inflationary trajectory}, and just present the final global phase portrait on figure \ref{fig:Starobinskyinfinite}. Here for the large field values the dynamics is dominated by the leading inflationary trajectory (orange) that starts from the asymptotic de Sitter fixed point $C_+$ and runs to the final attractor $A$ of the late universe. This trajectory ultimately collects the other trajectories that start from the other asymptotic source fixed point $D_+$, whereas the points $B$ and $E_+$ are saddles receiving and delivering only particular trajectories on the edge of the physical phase space. The leading trajectory undergoes a prolonged period of accelerated expansion (green zone), exits into decelerated expansion (white zone), and finally succumbs to residual damped oscillations around the point $A$, which are indicated by a dotted line. The slow roll approximation is marked by the dashed line, and similar to the case of the previous section, it predicts the end of the accelerated expansion regime earlier (higher field values) than experienced by the actual nonlinear solution. 

At the minimum of the potential ($\Phi=M$), the variables \eqref{eq:dynamical variables} stretch out the depiction of the phase space in the $z$ direction (red dotted curve at point $A$), meaning that any finite $\dot{\Phi}$ gets projected to the outer ring on the portrait, see the discussion at the end of Sec.\ \ref{subsec: CW phase portrait}. In fact, the field $\Phi$ is subject to oscillations around this value. Dedicated numerical investigations show that if we start with a sufficiently high $\dot{\Phi}$ at the exponential wing of the invariant potential (left from $A$), it is possible for the field to cross over to the plateau wing (right from $A$) and then even experience inflation by slowly rolling down close to the leading orange trajectory \cite{Mishra:2018dtg,Mishra:2019ymr}.

\begin{figure}[t]
	\centering
    \subfigure[]{
		\includegraphics[width=0.46\textwidth]{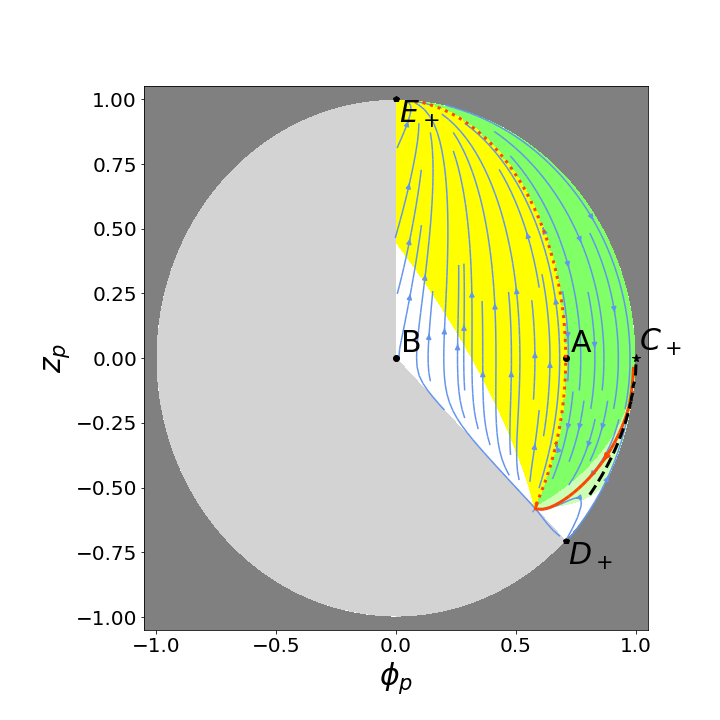} \label{fig:Starobinskyinfinite}}
	\subfigure[]{
		\includegraphics[width=0.46\textwidth]{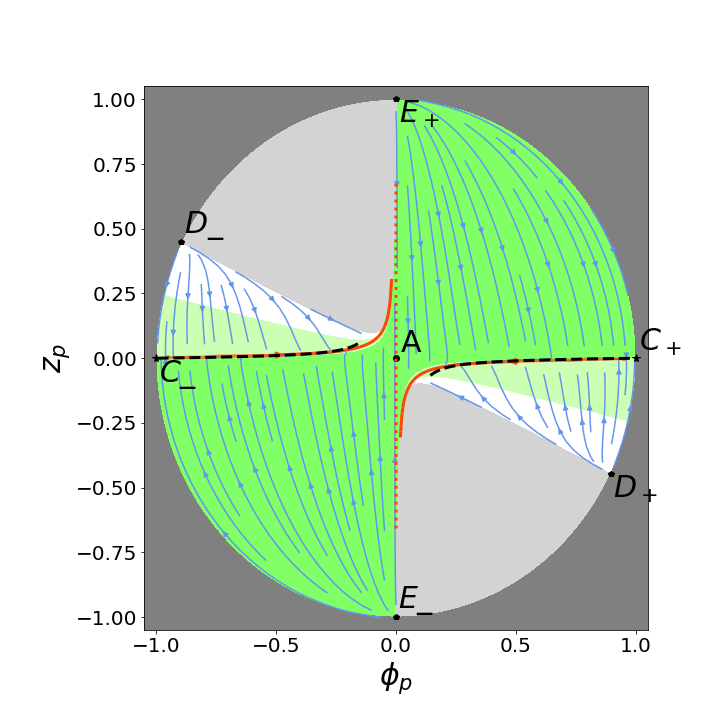} \label{fig:Higgsinfinite}}
  \\
    \subfigure[]{
		\includegraphics[width=0.46\textwidth]{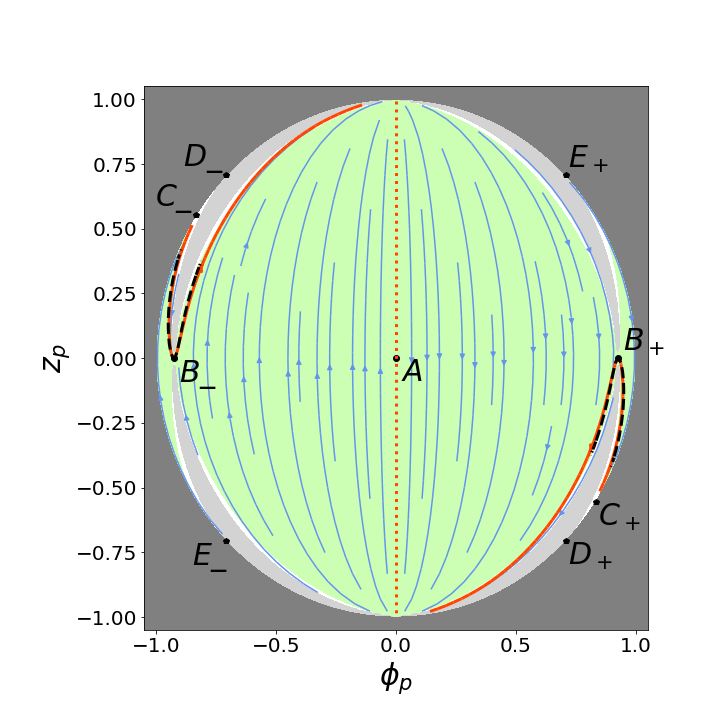} \label{fig:poleinfinite}}
    \subfigure[]{
		\includegraphics[width=0.46\textwidth]{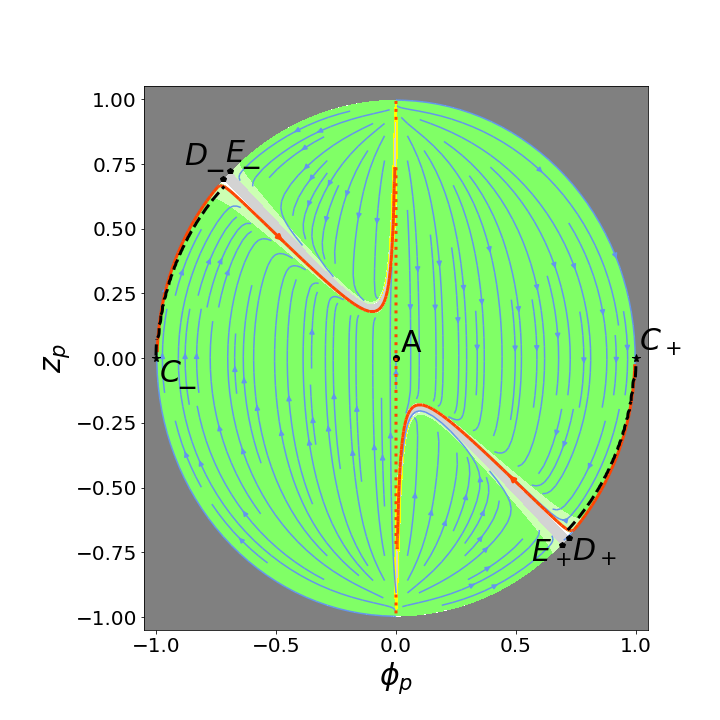} \label{fig:Palatininonminimalinfinite}}
  \caption{Cosmological phase portraits of observationally equivalent models: a) metric Starobinsky model \eqref{eq: Starobinsky metric model} with $\beta=40000$, b) metric Higgs model \eqref{eq: Higgs metric model} with $\xi=100$, $\lambda=0.129$, $v=0$, c) minimally coupled pole inflation model \eqref{eq: Pole metric model} with $\alpha=1$, $\lambda=10^{-4}$ d) nonminimally coupled Palatini model with $\xi=100$, $\lambda=10^{-5}$ and $\alpha=\sqrt{2/3}$ (in all cases $\MP=1$). The green background stands for superaccelerated, light green accelerated, white decelerated, and yellow superstiff expansion, while grey covers the unphysical region. Orange trajectories are heteroclinic orbits between the fixed points, and the dashed curve marks the path of slow roll approximation.}
\label{fig:Higgsequivalentplots}
\end{figure}

\subsection{Higgs inflation in metric formalism}
\label{higgsmetric}

Let us take the well known nonminimally coupled Higgs model in the metric formalism \cite{Bezrukov:2007ep},
 \begin{align}
 \label{eq: Higgs metric model}
    \mathcal{A} &= M^2 + \xi \Phi^2 \,, \qquad
    \mathcal{B} = 1 \,, \qquad
    \mathcal{V} = \frac{\lambda}{4} \left( \Phi^2 - v^2 \right)^2 \,,
\end{align}
where once again $M$ is a constant with dimensions of mass. 

In order to see the predicted observables, we write down the invariant field under the assumption that the expression \eqref{I_Phi} is dominated by the metric term, i.e.\ we are in the strong coupling limit that allows us to approximate
\begin{align}
    \IP \simeq \sqrt{\left(\frac{\mathcal{A}'}{\mathcal{A}}\right)^2}.
\end{align}
Solving this equation, we can write down the form of $\mathcal{A}$ directly in terms of $\IP$:
\begin{align}
    \mathcal{A} 
    \simeq M^2 e^{\sqrt{\frac{2}{3}} \IP} \,.
\end{align}
Therefore, using the explicit form of the coupling function $\mathcal{A} = M^2 +  \xi \Phi^2$, we can express the invariant field in terms of $\Phi$ as 
\begin{align}\label{eq: I_P metric Higgs}
  \IP \simeq \sqrt{\frac{3}{2}} \ln \left(1 +  \frac{\xi \Phi^2}{M^2}   \right).
\end{align}
Inverting this relation and substituting into \eqref{I_V} results in   the invariant potential
\begin{align}
\label{eq: I_V metric Higgs}
    \mathcal{I}_{\mathrm{V}} &= \frac{\lambda}{4 \xi^2} \left(1 + e^{-\sqrt{\tfrac{2}{3}} \mathcal{I}_\Phi} \right)^{-2} \simeq \frac{\lambda}{4 \xi^2} \left(1 - e^{-\sqrt{\tfrac{2}{3}} \mathcal{I}_\Phi} \right)^{2},
\end{align}
where in the last step we assumed $1 \ll \mathcal{I}_\Phi$. 

The mapping \eqref{eq: I_P metric Higgs} tells us that only the positive values of the invariant field are explored, and the flat plateau wing of the invariant potential characterizes the dynamics of both positive and negative $\Phi$. Given that the form of the potential \eqref{eq: I_V metric Higgs} is approximately coinciding with the corresponding expression \eqref{eq: I_V Starobinsky} of Starobisnky inflation, the algorithm of Ref.\ \cite{Jarv:2016sow} yields the same leading order prediction for the observables $n_s$ \eqref{eq: n_s Starobinsky} and $r$ \eqref{eq: r Starobinsky}.
The value of $M$ will affect only quantities with a conformal weight (i.e.\ those that do scale after a conformal transformation), such as the strength of the scalar spectrum. Taking its value to coincide with the reduced Planck mass (as the field settles in the VEV $v$ which approaches zero, recovering GR), we find the usual value of $\xi = 17000$ for the Standard Model quartic coupling $\lambda = 0.129$, in agreement with Ref.\ \cite{Bezrukov:2007ep}, and also consistent with our strong coupling assumption. By comparing the overall factors of the invariant potentials \eqref{eq: I_V Starobinsky} and \eqref{eq: I_V metric Higgs} we see that the observables $n_s$, $r$, and $A_s$ match for the Starobinsky and metric Higgs models in the leading order~if
$
    \frac{\lambda}{4 \xi^2} = \frac{1}{8\beta} \,.
$

Again, skipping all the calculational details, the global phase portrait of the model is presented on Fig.\ \ref{fig:Higgsinfinite}. In our variables, the portrait is independent of the Higgs self-coupling $\lambda$. However, the nonminimal coupling $\xi$ to gravity affects the positions of the asymptotic fixed points $D$ and $E$ and the boundaries of the physical phase space (see Ref.\ \cite{Jarv:2021qpp} for illustration). Despite that, all the qualitative features of the portrait including the leading inflationary trajectories from $C_\pm$ to $A$ (orange solid curves) and their slow roll approximations (black dashed lines) retain their presence for any $\xi>0$. On Fig.\ \ref{fig:Higgsinfinite} we have chosen $\xi=100$ which is not so large as demanded by $A_s$, but has the benefit of clearly showing the main features of the global phase portrait in a discernible manner (for larger $\xi$ see Ref.\ \cite{Jarv:2021qpp}). Comparing Figs.\ \ref{fig:Starobinskyinfinite} and \ref{fig:Higgsinfinite} we can note how qualitatively the Higgs phase space is a double copy of the large field Starobinsky phase space (the ``plateau'' wing of the invariant potential). Due to the mapping \eqref{eq: I_P metric Higgs} the equivalent of the 
negative $\IP$ Starobinsky phase space has no correspondence in the $\Phi$ variable,
and in reality there is no dynamical equivalent to the exponentially steep wing of the invariant potential in the Higgs case.

\subsection{Pole inflation}
\label{subsec: pole inflation}

Pole inflation involves the introduction of a pole in the kinetic term. This special form of the kinetic term drives inflation close to the pole rather than the potential (which instead sets the overall energy scale of inflation) \cite{Terada:2016nqg}. The popular $\alpha$-attractors feature a pole of order 2: the form of the potential is not particularly relevant as long as it is well-behaved around the pole, as the pole ``smooths it out,'' leading to convergent predictions \cite{Kallosh:2013yoa, Kallosh:2015lwa, Carrasco:2015pla}. For concreteness, we specify the action as having no further higher-order poles, which leads to: 
\begin{align}
\label{eq: Pole metric model}
    \mathcal{A} &= M^2 \,, \qquad
    \mathcal{B} = \frac{6 \alpha\, M^2 \, \Phi^2}{(6\alpha M^2- \Phi^2)^2} \,, \qquad
    \mathcal{V} = \frac{\lambda}{4} \Phi^4 \,.
\end{align}
Usually in the pole inflation literature the parameter $\alpha$ has mass dimension two,
but here we have separated out the dimensionful constant $M$ to make $\alpha$ dimensionless like in the other examples. It is natural to associate $M=\MP$. 
Here the poles are located at $\Phi = \pm \sqrt{6\alpha} M$. This model is identical in metric and Palatini, since the non-minimal coupling $\Acal$ is constant from the beginning, i.e.\ the scalar field couples minimally to curvature.

The invariant scalar field \eqref{I_Phi} is explicitly dimensionless,
\begin{align}
    \IP &= \left\{ \begin{array}{ll} \pm \sqrt{\frac{3\alpha}{2}} \ln \left(1-\frac{\Phi^2}{6\alpha M^2} \right) \,, & |\Phi| \leq \sqrt{6\alpha}M \,, \\
    \pm \sqrt{\frac{3\alpha}{2}} \ln \left(\frac{\Phi^2}{6\alpha M^2}-1 \right) \,, & |\Phi| \geq \sqrt{6\alpha}M \,. \end{array} \right. 
\end{align}
By choosing the ``$-$'' sign above such that we intuitively match positive values of the original field $\Phi$ within the poles to positive values of the invariant field, the invariant potential \eqref{I_V} is given by
\begin{align}
\label{eq: I_V pole inflation}
    \mathcal{I}_{\mathrm{V}} &= \left\{ \begin{array}{ll} \frac{\lambda}{4} \left( 1 - e^{-\sqrt{\frac{2}{3\alpha}} \mathcal{I}_\Phi} \right)^2 \,, & |\Phi| \leq \sqrt{6\alpha}M \,,  \\ 
    \frac{\lambda}{4} \left( 1 + e^{-\sqrt{\frac{2}{3\alpha}} \mathcal{I}_\Phi} \right)^2 \,, & |\Phi| \geq \sqrt{6\alpha}M \,. \end{array} \right. 
\end{align}
We can see that in the case where $\alpha=1$ and smaller field values we recover the particular example equivalent to the metric Starobinsky \eqref{eq: I_V Starobinsky} and metric Higgs \eqref{eq: I_V metric Higgs} invariant potentials, matching exactly when
$
     \frac{\lambda}{4} = \frac{1}{8 \beta} \,.
$
This ensures that the values of the observables $n_s$, $r$, and $A_s$ also agree in the first approximation level.

For general $\alpha$, the usual Starobinsky predictions are modified as:
\begin{align}
n_s = 1- \frac{2}{\widehat N},
\qquad
r = \frac{12 \alpha}{\widehat N^2}.
\end{align}
This reinforces the ``attractor'' nature of this class of models \cite{Galante:2014ifa}, since the value of $\alpha$ can be used to drive down the prediction for the tensor-to-scalar ratio. In general, pole inflation is controlled by the order and residue of the pole \cite{Terada:2016nqg}: the order sets $n_s$, which is why the second order poles are observationally favored, and both set $r$, which at the moment only imposes an upper bound in $\alpha$. This treatment further assumes close proximity to one of the poles, as well as the field being between the two poles, such that the invariant field evolves subject to a T-model-type potential: if these assumptions are violated and the field evolves outside of the poles, we are led to unusual cosmological scenarios that call for extending the potential in order for inflation to be realized~\cite{Karamitsos:2019vor}.

From the phase portrait \ref{fig:poleinfinite} we can see that from the poles $\Phi=\pm \sqrt{6\alpha}$ (points $B_\pm$) ``downwards'' to the origin $\Phi=0$ (point $A$) the model behaves qualitatively like plateau wing Starobinsky or nonminimal Higgs, i.e.\ from point $B$ exits an attractor trajectory which is initially characterized by slow roll, then enters the deceleration zone and later winds down to the late time attractor $A$. From the poles ``upwards'' where the invariant potential \eqref{eq: I_V pole inflation} has a different sign, the behavior is rather different and incompatible with the expectations for inflation. Namely, although there is an attractor trajectory from point $C$ to point $B$, it starts in a region of decelerated expansion, and only later when approaching point $B$ experiences slow roll and accelerated expansion without a graceful exit. For $\alpha<4/9$ the asymptotic de Sitter point $C$ would be located in the nonphysical region of the phase space and thus at large field values there would be no attractor trajectory to collect the other solutions which get drawn to the boundary $D \to B$ instead.

\subsection{Nonminimally coupled Palatini model}
\label{nonminPalatini}

We will now explore a few examples set in the Palatini formalism. To be really distinct from the metric models, the gravitational coupling $\Acal$ must not be constant but depend on the scalar field. In practice, finding Palatini models that give the same first order predictions about the spectrum as the Starobinsky and nonminimal Higgs models requires some fairly complicated actions. This complexity can manifest either in the gravitational coupling, the kinetic term or the potential. For example, consider the following model:
\begin{align}
\label{eq: nonminimal Palatini model}
\mathcal{A}&=M^2+ \xi \Phi^2 \,, \qquad 
\mathcal{B}=1 \,, \qquad 
\mathcal{V}=\lambda [M^2+\xi\Phi^2]^2 \left[ 1-
\left(\frac{M}{\sqrt{M^2+ \xi  \Phi ^2  } +  \sqrt{\xi}  |\Phi | } \right)^{\frac{\alpha}{\sqrt{\xi}}}
\, \right],
\end{align}
where $M$ again carries the dimension of mass, while the parameters $\xi$, $\lambda$, and $\alpha$ are positive and dimensionless. The complexity of this model appears in the form of an unusual non-integer power in the potential. The absolute value appearing in the potential ensures that it remains positive for all values.

We can find the invariants in terms of the original field as follows: 
\begin{align}
\label{eq: nonminimal Palatini I_Phi}
\mathcal{I}_\Phi &= \frac{1}{\sqrt{\xi }}{\rm atanh}\left(\frac{\sqrt{\xi } \Phi }{\sqrt{M^2+\xi  \Phi ^2}}\right),
\\
\IV &= \lambda   \left[ 1-
\left(\frac{M}{\sqrt{M^2+ \xi  \Phi ^2  } +  \sqrt{\xi}  |\Phi | } \right)^{\frac{\alpha}{\sqrt{\xi}}}\, \right].
\end{align}
We could solve for $N(\Phi)$, but it suffices to show that the invariant potential is Starobinsky-like. Indeed, we find that the invariant potential in terms of the invariant field is 
\begin{align}
\label{eq: nonminimal Palatini I_V}
\IV &=
\lambda \left[ 1-e^{-\alpha  {\cal I}_\Phi }\right].
\end{align}
This is a model of chaotic inflation \cite{Goncharov:1983mw} that nonetheless is similar to Starobinsky-like inflation in that it features a plateau:
\begin{align}
\IV &\simeq \lambda \left[ 1-e^{-\alpha  {\cal I}_\Phi - \ln\frac{1}{2} }\right]^2 \,.
\end{align}
Here we made the approximation for a large exponent, meaning that the additive term in the exponent is also insignificant. The invariant potential \eqref{eq: nonminimal Palatini I_V} matches the Starobinsky model invariant potential \eqref{eq: I_V Starobinsky} if $\alpha=\sqrt{2/3}$ and
$
    \lambda = \frac{1}{8 \beta} \,.
$
As a result, the respective Starobinsky model observables are recovered once again.

The corresponding global phase portrait is given on Fig.\ \ref{fig:Palatininonminimalinfinite}. The invariant potential has a minimum at $\IP=0$ which by \eqref{eq: nonminimal Palatini I_Phi} maps to $\Phi=0$ and the minimum of the original potential \eqref{eq: nonminimal Palatini model} also. It corresponds to the point $A$ which is the final destination of scalar field dynamics. We can set $M=\MP$ as late time universe would have $\Acal\approx M^2$ and show general relativity like behavior. Qualitatively the model \eqref{eq: nonminimal Palatini model} has similar features to the plateau arm part of the Starobinsky model as well as the nonminimal metric Higgs model, whereby the leading inflationary trajectory starts at an asymptotic de Sitter point $C$ and runs into $A$, while other trajectories start at $D$ and approach the course of the leading trajectory. The mapping \eqref{eq: nonminimal Palatini I_Phi} tells that slow roll behavior takes place only at very high $\Phi$ values (in late time $\MP$ units), like in the Starobinsky case and unlike the nonminimal Higgs case. Similar to the induced gravity Coleman-Weinberg Palatini model (Fig.\ \ref{fig:PalatiniCWinfinite}, there seems to be a pocket of physically allowed initial conditions which allow superacceleration, but without a connection to the slow roll inflation.

\subsection{Higgs inflation with noncanonical kinetic term in the Palatini formalism}
\label{subsec: Higgs noncanonical Palatini}

Another option in Palatini is to consider models whose complexity is made manifest in the kinetic term. Such models feature a vanishing kinetic term at large field values, but unlike pole inflation, there does not have to be a divergence at select points. Consider the following action describing a non-canonical model:
\begin{align}
\label{eq: Noncanonical Higgs Palatini model}
    \mathcal{A} &= M^2+\xi \Phi^2 \,, \qquad
\mathcal{B} = \frac{6 \alpha M^4 \Phi^2}{(M^2+\xi\Phi^2)(M^2+(\xi-1)\Phi^2)^2} \,,
\qquad
    \mathcal{V} = \frac{\lambda}{4} \Phi^4 \,.
\end{align}
As in the previous examples, $M$ has the dimension of mass, while the parameters $\xi$, $\lambda$, and $\alpha$ are positive and dimensionless. The kinetic term features a pole if $\xi<1$.

Remembering that we are in the Palatini formalism and assuming $\xi>1$
\begin{align}
\label{eq: I_Phi higgs palatini noncanonical inflation}
\mathcal{I}_{\mathrm{\Phi}} &= 
\sqrt{\frac{3\alpha}{2}}  \ln \left(\frac{M^2+(\xi -1) \Phi ^2}{M^2+\xi  \Phi ^2}\right) \,.
\end{align}
Inverting and substituting as usual, we obtain the invariant potential \eqref{I_V}
\begin{align}
\label{eq: Iv Palatini noncanonical}
\IV &= \frac{\lambda }{\left[1+\coth \left(\frac{{\cal I}_{\rm \Phi} }{  \sqrt{6\alpha }}\right)\right]^2} \simeq \frac{\lambda}{4} \left(1 - e^{-\sqrt{\tfrac{2}{3 \alpha}} \mathcal{I}_\Phi} \right)^{2},
\end{align}
where in the last step large $\IP$ was assumed. It occurs that in the invariant potential there is no dependence on $\xi$, which has been absorbed into the invariant field. The invariant potentials \eqref{eq: I_V Starobinsky} and \eqref{eq: Iv Palatini noncanonical} match exactly when $\alpha=1$ and
$
    \frac{\lambda}{4} = \frac{1}{8 \beta} \,.
$
In that case, the values of the observables $n_s$, r, and $A_s$ would also agree in the first approximation level to the predictions of the Starobinsky model, Sec.\ \ref{subsec: metric Starobinsky}.

Looking more carefully at Eq.\ \eqref{eq: I_Phi higgs palatini noncanonical inflation}, we notice that for $\xi>1$, only negative values of the invariant field $\IP$ correspond to the original field $\Phi$. 
In this case, we always find ourselves on the exponentially steep side of the invariant potential which is not conducive to inflation, not in the desirable plateau region. Although it is still possible to draw the phase portrait, Fig.\ \ref{fig:PalatininoncanonicalHiggsinfinite}, it does not exhibit a slow roll curve, since that regime is not available. Nevertheless, it is interesting to see that the Friedmann constraint \eqref{eq:Friedmann_bound} is satisfied for all values of the variables and there are no restrictions on the physical phase space. Some features of the dynamics are still recognizable and similar to the previous cases, like the existence of asymptotic de Sitter fixed point $C$, the final attractor point $A$, and a master trajectory from $C$ to $A$ which attracts all other solutions. Nonetheless, the dynamics exhibited by this scenario are superacceleration-type without an end. A reasonable choice would be to set $M=\MP$ to recover $\Acal \to \MP^2$ when the scalar field settles down, $\IP \to 0$ and $\Phi \to 0$.

In contrast, for $\xi<1$ poles are introduced to the kinetic coupling $\Bcal$. More precisely, after conformally eliminating the nonminimal coupling for $0<\xi<1$ there are two second order poles at $\pm M/\sqrt{1-\xi }$, which break the field space into three disconnected subspaces and the situation is somewhat similar to the minimal pole inflation case of Sec.\ \ref{subsec: pole inflation}, Fig.\ \ref{fig:poleinfinite}. If we considered $\xi<0$ then two additional second order poles are introduced at $\pm M/\sqrt{-\xi}$. Since the field cannot cross poles, the field space is now broken up into five disconnected subspaces, and viable inflation occurs only near the poles.

\begin{figure}[t]
	\centering
    \subfigure[]{
		\includegraphics[width=0.46\textwidth]{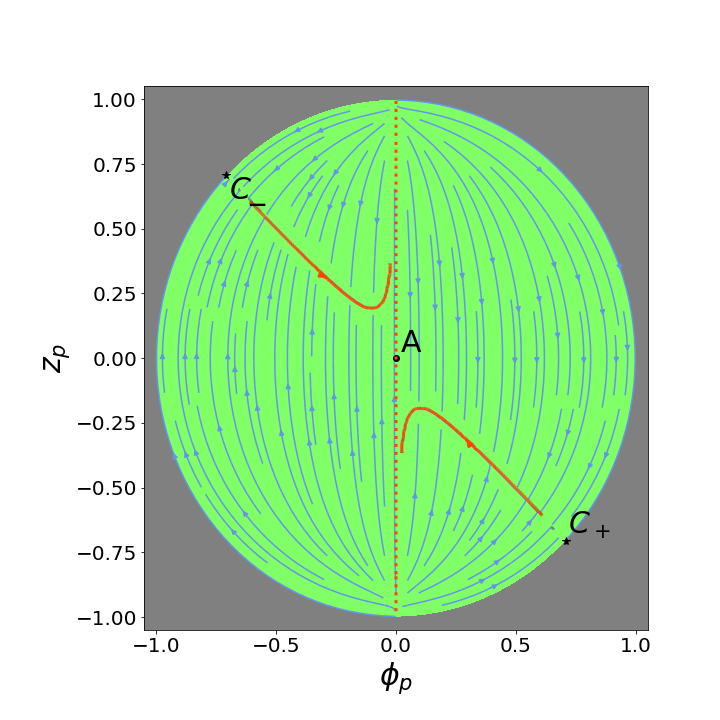} \label{fig:PalatininoncanonicalHiggsinfinite}}
	\subfigure[]{
		\includegraphics[width=0.46\textwidth]{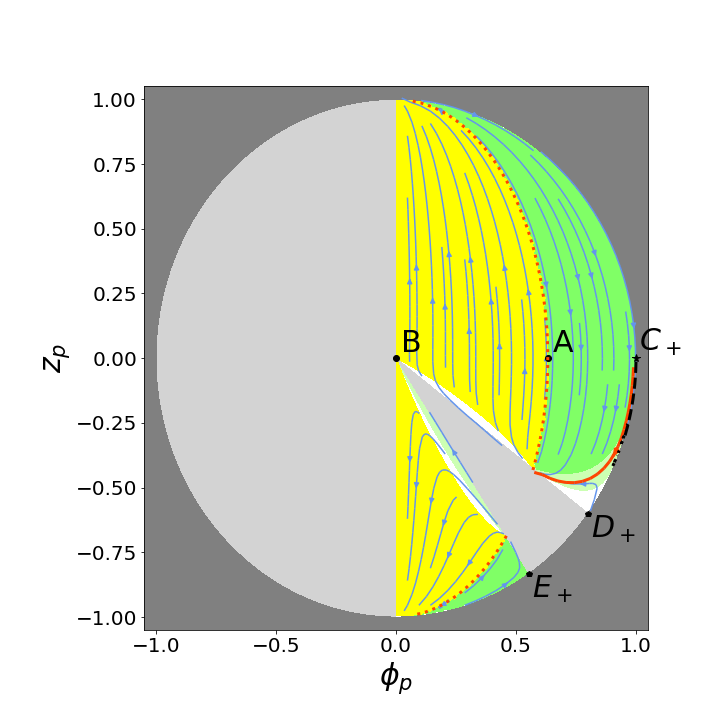} \label{fig:inducedPalatiniinfinite}}
  \caption{Continuation of cosmological phase portraits of observationally equivalent models: a) Palatini noncanonical Higgs model \eqref{eq: Noncanonical Higgs Palatini model} with $\alpha=1$, $\lambda=1.29\times 10^{-5}$ b) Palatini induced gravity model \eqref{eq: induced gravity Palatini model} with $\lambda=9.5\times 10^{-11}$, $n=2/3$, $\Phi_0=\sqrt{2/3}$. Green background stands for superaccelerated, light green accelerated, white decelerated, and yellow superstiff expansion, while grey covers the unphysical region. Orange trajectories are heteroclinic orbits between the fixed points, and the dashed curve marks the path of slow roll approximation.}
\label{fig:Higgsequivalentplots2}
\end{figure}

\subsection{Induced gravity Palatini model}
\label{inducedpalatini}
Finally, we examine an induced gravity-like model, where there is no mass scale $M$ in the non-minimal coupling. In this case, we take Palatini formalism and specify the action by the following model functions (compare with example V.A.2 in Ref.\ \cite{Jarv:2020qqm}):
\begin{align}
\label{eq: induced gravity Palatini model}
\mathcal{A}&=\frac{3}{2} \Phi^2 \,, \qquad 
\mathcal{B}=1 \,, \qquad 
\mathcal{V}= \frac{9}{4} \lambda M^{2n} \Phi^4 \left(\Phi_0^{-n}-\Phi^{- n} \right)^2 \,,
\end{align}
where $n$ and $\lambda$ are positive parameters, while $\Phi$ and $\Phi_0$ are assumed to be nonnegative to make sure the action is real for any $n$.
The invariant field \eqref{I_Phi} and potential \eqref{I_m} can be found as follows: 
\begin{align}
\IP &= \sqrt{\frac{2}{3}}   \ln \left(\frac{\Phi}{\Phi_0} \right),\\
\IV &= \lambda  M^{2n} \left(\Phi_0^{-n}-\Phi ^{-n}\right)^2,
\end{align}
while in terms of the invariant field the invariant potential is given by
\begin{align}
\label{eq: induced gravity Palatini IV}
\IV = \lambda \frac{M^{2n}}{\Phi_0^{2n}} \left(1 -e^{-n\sqrt{\frac{3}{2}} \IP} \right)^2 \,.
\end{align}
Comparing with the Starobinsky invariant potential \eqref{eq: I_V Starobinsky} we see a match when $n=\frac{2}{3}$ and
\begin{align}
    \lambda \left( \frac{M}{\Phi_0}\right)^{\frac{2}{3}} &= \frac{1}{8 \beta} \,.
\end{align}
In analogy with Sec.\ \ref{subsec: metric Starobinsky} positive invariant field represents large original field, $\Phi_0 < \Phi$, and the plateau like invariant potential allows inflationary dynamics. Negative invariant field values represent a small original field, $0<\Phi<\Phi_0$, where exponentially growing invariant potential is generally not conducive to slow roll and inflation. The attractive fixed point of the late universe is at $\IP=0$, i.e.\ $\Phi=\Phi_0$. It makes sense to set $\Phi_0=\sqrt{2/3}\MP$ since then at late times $\Acal\to \MP^2$ like in general relativity.
The phase portrait is given on Fig. \ref{fig:inducedPalatiniinfinite}. It is in many ways similar to the Starobinsky case on Fig.\ \ref{fig:Starobinskyinfinite}, except there is a pocket of physically allowed phase space completely disconnected from the remainder where the inflationary trajectory occurs.

\section{Identical actions that yield different observational parameters in metric and Palatini}
\label{sec:identicalactionsdifferentobs}

In this section, we will study the discrepancy arising in observables and the change in the global dynamics if we switch between metric and Palatini. We will study two models we have previously looked at, except under a different approach, before comparing the two.

\subsection{Higgs inflation in Palatini formulation}

We have already studied Higgs inflation in the metric formalism in Subsection~\ref{higgsmetric}. We can now study the model \eqref{eq: Higgs metric model} in the Palatini approach \cite{Bauer:2008zj,Rasanen:2017ivk}. We find that the invariant quantities are
\begin{align}
\label{eq: Higgs Palatini IP}
   {\cal I}_{\Phi} &= 
\frac{1}{\sqrt{\xi }}{\rm atanh}\left(\frac{\sqrt{\xi } \Phi }{\sqrt{M^2+\xi  \Phi ^2}}\right),
  \\
    {\cal I}_{V} &= 
    \frac{\lambda \Phi^4}{(M^2+ \xi \Phi^2)^2} \,.
\end{align}
Writing the potential in terms of the invariant field gives now a different function
\begin{align}
  {\cal I}_{V} = \frac{\lambda}{\xi^2} \tanh^4{(\sqrt{\xi} \ \IP)} .
\end{align}
In the case where $\IP$ and $\IV$ are the minimally coupled field and potential, this type of model is known as a T-model in the literature \cite{Kallosh:2013hoa, Sloan:2019jyl}.
The invariant potential still features an asymptotic plateau like in the metric case. Using the potential, it is possible to calculate the observables \eqref{ns} and \eqref{r} as 
\begin{align}
n_s = 1 -\frac{2}{\widehat{N}} 
\qquad
r = \frac{2}{\xi  \widehat{N}^2}.
\end{align}
As the metric Higgs observables coincided to the first order with the observables of the Starobinsky model \eqref{eq: n_s Starobinsky}, \eqref{eq: r Starobinsky}, we note that for the parameter value $\xi = 1/6$ the Palatini Higgs model predicts an identical result to metric Higgs. However in general, the metric and Palatini Higgs differ, since $\xi$, unlike in the metric case, does not appear only as a prefactor of the invariant potential and affects the tensor-to-scalar ratio $r$. On the other hand, matching with the amplitude of the spectrum returns $\xi \sim 10^5$ to $10^9$ \cite{Gialamas:2020vto}, meaning that the model does diverge between metric and Palatini. However, the final state $\Phi=0$ is still the same as in the metric case, and we can set $M=\MP$.

\begin{figure*}
	\centering
	\subfigure[]{
		\includegraphics[width=0.45\textwidth]{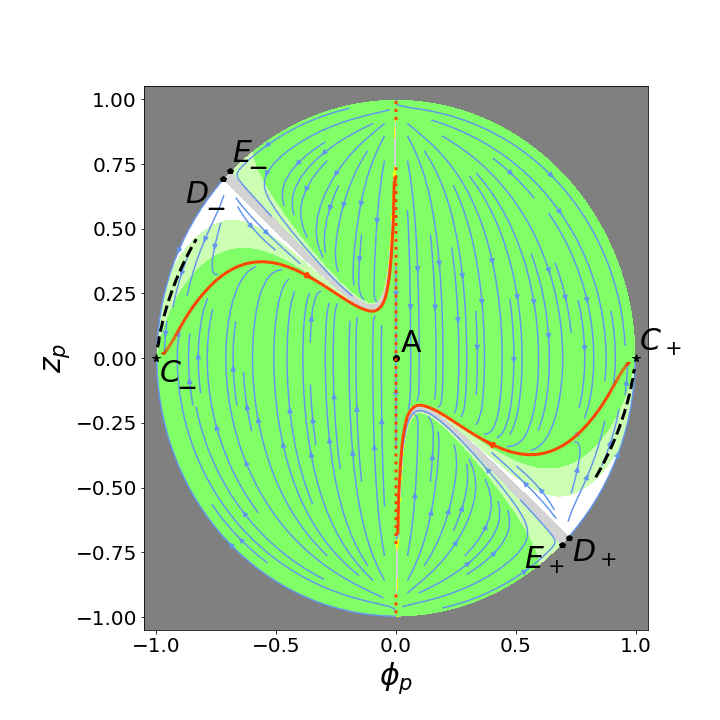} \label{fig:HiggsPalatiniinfinite}}
  \subfigure[]{
		\includegraphics[width=0.45\textwidth]{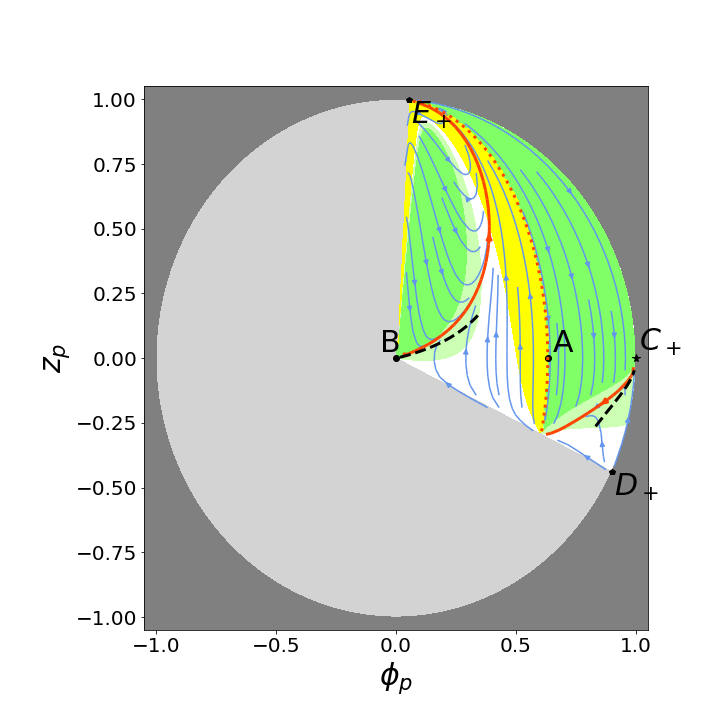} \label{fig:inducedgravitymetricinfinite}
}
	\caption{Cosmological phase portraits of a) the Palatini Higgs model \eqref{eq: Higgs metric model} with $\xi=100$, $\lambda=0.129$, $\Lambda=0$ and b) induced gravity in the metric formalism model with $\lambda=9.5\times10^{-11}$, $n=2/3$, $\Phi_0=\sqrt{2/3}$ . Green background stands for superaccelerated, light green accelerated, white decelerated, and yellow superstiff expansion, while grey covers the unphysical region. Orange trajectories are heteroclinic orbits between the fixed points, and the dashed curve marks the path of slow roll approximation.}
\label{fig:Nonequivalentplots}
\end{figure*}

The corresponding phase portrait Fig. \ref{fig:HiggsPalatiniinfinite} is drawn for the same parameter values as for the nonminimal Higgs model in the metric formalism, Fig.\ \ref{fig:Higgsinfinite}. Some features of these pictures are qualitatively the same, like the leading inflationary trajectory $C \to A$ approximated by a slow roll curve, and the existence of the other asymptotic fixed points $D$ and $E$. However, the extent of the physically allowed phase space is different whereby an extra ``pocket'' emerges in comparison to the metric case, and the points $D$ and $E$ correspond to different expansion regimes, similarly to the more detailed discussion in the Coleman-Weinberg example (Sec.\ \ref{subsec: CW phase portrait}). One can also not notice that the scalar field value at the end of inflation, $\Phi^*$, predicted by the slow roll approximation and the full nonlinear solution differ from each other rather remarkably in the Palatini case.

\subsection{Induced gravity inflation in metric formulation}
Finally, let us study the metric counterpart to induced gravity inflation considered in Subsection~\ref{inducedpalatini}. For the model functions~\eqref{eq: induced gravity Palatini model} under the metric formalism, the invariant quantities are now given as 
\begin{align}
\mathcal{I}_\Phi &= 2\sqrt{\frac{5}{3}}   \ln \left(\frac{\Phi}{\Phi_0}\right), \\
\IV &=  \lambda M^{2n} \left(\Phi_0^{-n}-\Phi ^{-n}\right)^2,
\end{align}
where the invariant potential in terms of the invariant field is
\begin{align}
{\cal I}_V = \lambda \frac{M^{2n}}{\Phi_0^{2n}} \left(1 -e^{-n\sqrt{\frac{3}{20}} {\cal I}_\Phi} \right)^2 
\end{align}
This is still a Starobinsky-like potential, only with the parameter $n$ multiplied by a slightly different factor in the exponential, when compared to the Palatini counterpart \eqref{eq: induced gravity Palatini IV}. It would match the Starobinsky model observables when $n=\sqrt{40/9}$ and
\begin{align}
    \lambda \left( \frac{M}{\Phi_0}\right)^{\sqrt{\frac{40}{9}}} &= \frac{1}{8 \beta} \,.
\end{align}. 

However for comparison, on Fig.\ \ref{fig:inducedgravitymetricinfinite} we use the same parameter values as on Fig.\ \ref{fig:inducedPalatiniinfinite} for the the metric case. For large field values where both the metric and Palatini models are described by the plateau wing of the Starobinsky-like invariant potential, the dynamics is rather similar. In both cases, there is a leading inflationary trajectory $C_+ \to A$ approximated by a slow roll curve. The final attractor point $A$ corresponds to the minimum of the potential, $\Phi=\Phi_0$, i.e.\ $\IP=0$. For small field values, we see a stretch of a superstiff expansion region similar to the one Fig.\ \ref{fig:inducedPalatiniinfinite} left of the point $A$. However, as we move closer to the point $B$ at $\Phi=0$, the possibility for accelerated expansion and slow roll emerges, in contrast to the Palatini case. This may come as a surprise initially, but it is not in contradiction with the earlier statements about the general congruence of scalar field dynamics between metric and Palatini. First of all, strictly speaking, the limit $\Phi \to 0$ is not included in the assumptions \eqref{eq: assumptions on the model functions} of the present study, since in said limit, the nonminimal coupling function $\Acal(\Phi)$ vanishes and gravity disappears entirely. Without a deeper investigation, we cannot claim that the correspondence of the fixed point structure still holds there. Second, the overall dynamics of the scalar field in the region is still qualitatively congruent to the Palatini case, as the field consistently evolves from $B$ to $A$. However, since the effective barotropic index \eqref{eq:weff} is sensitive to the choice between metric and Palatini, the expansion properties are different, including whether slow roll is achieved or not. Visually, the metric version of the induced gravity model \eqref{eq: induced gravity Palatini model} on Fig.\ \ref{fig:inducedgravitymetricinfinite} bears similarities to the quintic potential induced gravity model \eqref{eq: CW model} studied in Sec.\ \ref{sec:identicalactionsidenticalobs}, Fig.\ \ref{fig:CWmetricandPalatini}, but without a dedicated analysis and careful attention to the singular limits, it is difficult to make more rigorous statements.

\section{Conclusions}
\label{sec:conclusions}

In this paper, we have applied the techniques of Ref.\ \cite{Jarv:2021qpp} to study the global phase portraits of different single scalar field inflationary models, both in the metric and Palatini formalisms defined in the Jordan frame. As a first step, we derived the general equations as well as the flat FLRW cosmological equations for a scalar field nonminimally coupled to the Ricci curvature scalar in a unified framework that encompasses both the metric and the Palatini formalisms in the Jordan frame. So far, the Palatini scalar-tensor gravity has been mostly studied in the Einstein frame, and the Jordan frame description is scarce to come across in the literature. We characterized the cosmological evolution by the notions of effective potential and effective mass, and showed that despite different terms present in the underlying equations the scalar field fixed points and their key characteristics coincide in the metric and Palatini cases. Moreover, we derived the expression for the scalar field slow roll regime, including a novel expression in the Palatini formalism, and extended (from metric to Palatini) the earlier observation of the evolution of the slow roll curve from a de Sitter or asymptotically de Sitter fixed point.

Using the effective algorithm of Refs.\ \cite{Jarv:2016sow,Jarv:2020qqm} to compute the spectrum of perturbations generated by the inflation, it is easy to check that the same action can in some special cases give identical predictions for the spectral index $n_s$ and tensor to scalar ratio $r$ in the metric and Palatini formalisms, but in general these predictions are different. On the other hand, there can be several rather different actions studied either under the metric or the Palatini formalism that lead to the same observational predictions. This gives rise to a degeneracy among the models. In order to shed light beyond that degeneracy, we turned to the global phase portraits, which are a means to get a complete picture of all solutions that correspond to a given model, which helps to identify the asymptotic states and general qualitative features of the dynamics. Thus, in Sec.\ \ref{sec:identicalactionsidenticalobs}, we compared in detail a case of the same action giving the same observables in metric and Palatini, taking the example of induced gravity and quintic potential with Coleman-Weinberg correction. Next, in Sec.\ \ref{sec:differentactionsidenticalobs}, we looked at several models with different actions that give the same values for the observables, starting with the famous Starobinsky and metric Higgs models, and then adding to the mix a minimally coupled second order pole model, as well as three Palatini models with specific model functions. Finally, in Sec.\ \ref{sec:identicalactionsdifferentobs}, we took two examples from the previous section and switched between metric and Palatini, in order to illustrate the possibility of the same action giving rise to different observables depending on the choice of formalism.

Overall, we see that models which in some limit are described by the same invariant potential also exhibit a qualitatively similar phase space for the original scalar field in the same limit. It is important to realize that inflation in the phase space is not characterized by a single fixed point, but rather by a trajectory from a saddle-type fixed point (on the plots marked by~$B$ or~$C$) to an attractor-type fixed point (marked by $A$). This trajectory collects other nearby trajectories which end up following it: that explains the robustness of inflation with respect to initial conditions. The saddle-type fixed point may reside in the bulk of the phase space ($B$) or in the asymptotics ($C$). Since the fixed point structure is the same in the metric and Palatini cases, the viability of slow-roll evolution and inflation for a given action is shared in both formalisms, although the resultant evolution and observables are not necessarily the same. It is true that different actions that give rise to the same invariant potentials and predict the same observables also have similar corresponding fixed points and phase space features for the original field. However, before jumping to conclusions based on the invariant potential, one should be vigilant and make sure that the original field does indeed map to the part of the invariant potential that supports inflation, otherwise an inflationary period may never occur, as we saw in Sec.\ \ref{subsec: Higgs noncanonical Palatini}.

Differences among the models that predict the same observables become manifest when we ``zoom out'' from the leading inflationary trajectory and look at the global picture. First of all, the limits of the physically available phase space, i.e.\ where the Friedmann constraint equation is satisfied, will be typically different. Perhaps the most evident illustration of this aspect is seen in the separate ``pockets'' of the phase space from where the initial conditions do not seem to give a chance to inflation, i.e.\ the trajectories there do not come close to the leading inflationary orbit later, in contrast to the other parts of the phase space where the inflation orbit is present and dominant. In the examples we considered, such ``pockets'' occurred only in the Palatini models but not in the metric models. At the moment it is hard to tell whether that feature is something typical for the Palatini case, or just a property of the particular examples we looked at. In principle it might also happen that a fixed point which resides in a physical phase space in one formalism, might find itself in a nonphysical part of the phase space in another formalism, thus completely altering the dynamics (see e.g. \cite{Jarv:2021ehj}).

Another feature that is different between almost any model (regardless of whether predictions are convergent or not) is the quantitative dynamics of the past asymptotic state of the system. These dynamics are represented by the points $D$, which act as sources for a large bundle of trajectories in the phase space (and in many cases practically all of them). In terms of the global phase space variables we use, it is possible to determine the coordinates of that point and then compute the respective evolution of the scalar field and scale factor in cosmic time. 

In Sec.\ \ref{subsec: CW inflationary trajectory}, we also discussed in passing the different possible definitions of the end of inflation and the amount of expansion in terms of e-folds. The exit from acceleration slightly differs in the Jordan and invariant (Einstein) variables, as does the count in the number of e-folds, which can affect the precise predictions of the spectrum of perturbations and the modelling of the subsequent particle creation. We also confirm the earlier observations \cite{Urena-Lopez:2007zal,Grain:2017dqa,Jarv:2021qpp} that the slow roll approximation puts the end of inflation to a scalar field value that is too early, as the actual nonlinear master solution experiences accelerated expansion regime a while longer to scalar field values further on. In some cases that discrepancy can be rather significant. In our examples, the mismatch between the slow roll and the full nonlinear solution is especially pronounced in the Palatini models, but it is unclear whether this is a generic feature or an artifact of the particular models under consideration. 

The method of dynamical systems and the presentation of phase diagrams have has provided a great toolbox to analyze and visualize the properties of the multitude of models addressing late time universe. These tools are available to the study of the early universe as well, and as we have seen, can be invoked to gain new insights and a deeper understanding of the models of inflation. Several problems can be immediately posed for further investigation. One such question is how other schemes beyond the traditional slow roll scenario feature in the phase space: these include fast roll \cite{Linde:2001ae}, rapid roll \cite{Kofman:2007tr}, ultra-slow roll \cite{Martin:2012pe}, constant roll \cite{Motohashi:2014ppa}, or otherwise non-slow-roll \cite{Tasinato:2020vdk}. Furthermore, what are the phase space reasons for why certain types of models are attractors (e.g.\ $\alpha$-attractors and $\xi$-attractors) in the metric formalism \cite{Kallosh:2013yoa,Galante:2014ifa} but not in Palatini \cite{Jarv:2017azx}?  How are the global properties of the phase space affected, when one tries to take into account various quantum corrections \cite{DeSimone:2008ei,Finn:2019aip}, or when the inflaton field is nonminimally coupled to other gravitational terms like Gauss-Bonnet \cite{Carter:2005fu} or other higher curvature corrections \cite{Satoh:2008ck}, or as in the Horndeski theory \cite{Kobayashi:2011nu}?  Finally, do the same key dynamical properties remain when one changes the underlying formalism to more general metric-affine settings \cite{Shaposhnikov:2020gts,Rigouzzo:2022yan}? In the long run, the study of global phase portraits for different formalisms may make a ``dictionary'' between various models possible, allowing for a deeper understanding of the relation between their dynamics and the domains from which approximate physical predictions can be extracted.

\subsection*{Acknowledgements} LJ and MS were supported by the European Regional Development Fund through the Center of Excellence TK133 ``The Dark Side of the Universe" and by the Estonian Research Council grants PRG356 ``Gauge Gravity'' and TK202 ``Foundations of the Universe''. SK would like to thank the Department of Nuclear and Particle Physics at the National and Kapodistrian University of Athens for their hospitality where part of this work was completed.
 
\bibliographystyle{utphys}
\bibliography{scalartensor_tidied}


\end{document}